\documentclass[11pt]{article}

\usepackage[final]{acl}

\usepackage{times}
\usepackage{latexsym}

\usepackage[T1]{fontenc}
\usepackage[utf8]{inputenc}

\usepackage{microtype}

\usepackage{inconsolata}

\usepackage{graphicx}
\usepackage{subcaption}
\usepackage{booktabs}      % 提供 \toprule, \midrule, \bottomrule 命令
\usepackage{caption}       % 提供 \captionsetup 命令
\usepackage[dvipsnames]{xcolor}
\usepackage[table]{xcolor}
\usepackage{amsmath}       % 用于数学符号，如箭头
\usepackage{natbib}        % 用于 \citep 命令 (这是一个占位符，您需要有 .bib 文件)
\usepackage{amssymb}

\usepackage{arydshln}
\usepackage{fontawesome}
\definecolor{lightpurple}{RGB}{248, 243, 206} % 定义 lightpurple 颜色
\definecolor{darkgreen}{rgb}{0.0, 0.5, 0.0}   % 定义一个深绿色，更易读
\definecolor{darkred}{rgb}{0.8, 0.0, 0.0}     % 定义一个深红色
\usepackage{multirow}
\usepackage{makecell}
\usepackage{enumerate}
\usepackage{enumitem}
\usepackage{pifont}
\usepackage{graphicx}
\usepackage{bbding}
\usepackage{algorithm}          % algorithm float
\usepackage{algpseudocode}      % optional, for pseudo‑code
\usepackage{array}              % for p{…} column specifiers

\newcommand{\note}[1]{{\footnotesize\color{blue}{#1}}}
\usepackage[table,dvipsnames]{xcolor}
\usepackage{colortbl}
\usepackage{xspace}

\definecolor{ForestGreen}{RGB}{34,139,34}
\definecolor{myyellow}{RGB}{181, 181, 27}

\newcommand{\blue}[1]{$_{\color{BlueGreen}\downarrow #1}$}
\newcommand{\red}[1]{$_{\color{RedOrange}\uparrow #1}$}
\definecolor{darksalmon}{rgb}{0.91, 0.59, 0.48}
\definecolor{emerald}{rgb}{0.31, 0.78, 0.47}
\definecolor{greenpigment}{rgb}{0.0, 0.65, 0.31}
\definecolor{amaranth}{rgb}{0.9, 0.17, 0.31}

\definecolor{iris}{rgb}{0.35, 0.31, 0.81}
\definecolor{uu}{rgb}{0.95, 0.51, 0.51}
\definecolor{spirodiscoball}{rgb}{0.06, 0.75, 0.99}

\title{TopoDIM: One-shot Topology Generation of Diverse Interaction Modes for Multi-Agent Systems}

\author{
  \textbf{Rui Sun}$^{1}$ \quad
  \textbf{Jie Ding}$^{1}$ \quad
  \textbf{Chenghua Gong}$^{1}$ \quad
  \textbf{Tianjun Gu}$^{2}$ \\
  \textbf{Yihang Jiang}$^{1}$ \quad
  \textbf{Juyuan Zhang}$^{1}$ \quad
  \textbf{Liming Pan}$^{1{*}}$ \quad
  \textbf{Linyuan Lü}$^{1{*}}$
  \\
  $^{1}$University of Science and Technology of China, Hefei, China \\
  $^{2}$East China Normal University, Shanghai, China \\
  \texttt{\{rrsun,jieding25,gongchenghua,jiang\_yihang,zhangjuyuan2020\}@mail.ustc.edu.cn},\\
  \texttt{51275901043@stu.ecnu.edu.cn}, \texttt{\{pan\_liming, linyuan.lv\}@ustc.edu.cn}
}
\begin{document}
\maketitle
\begingroup
\renewcommand\thefootnote{}
\footnotetext{{*} Corresponding Author.}
\endgroup
\begin{abstract}
  Optimizing communication topology in LLM–based multi-agent system is critical for enabling collective intelligence.
  Existing methods mainly rely on spatio-temporal interaction paradigms, where the sequential execution of multi-round dialogues incurs high latency and computation.
  Motivated by the recent insights that evaluation and debate mechanisms can improve problem-solving in multi-agent systems, we propose \textsc{\textbf{TopoDIM}}, a framework for one-shot \textsc{\textbf{Topo}}logy generation with \textbf{D}iverse \textbf{I}nteraction \textbf{M}odes.
  Designed for decentralized execution to enhance adaptability and privacy, \textsc{TopoDIM} enables agents to autonomously construct heterogeneous communication without iterative coordination, achieving token efficiency and improved task performance.
  Experiments demonstrate that \textsc{TopoDIM} reduces total token consumption by 46.41\% while improving average performance by 1.50\% over state-of-the-art methods.
  Moreover, the framework exhibits strong adaptability in organizing communication among heterogeneous agents.
  Code is available at: \url{https://github.com/Sundiasy/TopoDIM}.
\end{abstract}
\section{Introduction}

Large language model (LLM)–based multi-agent systems (MAS) exhibit collective intelligence that can surpass the capabilities of a single LLM, achieving strong performance in mathematical computation \cite{lei2024macm}, code generation \cite{islam2024mapcoder}, software development \cite{he2025llm}, and scientific discovery \cite{ghareeb2025robin}. Prior studies \cite{zhuge2024gptswarm, zhang2025g, li2025assemble} have demonstrated that carefully designed interaction topologies enhance the task-processing capability of MAS through effective collaboration.

% Multi-agent systems (MAS) based on large language models (LLMs), which exhibit emergent collective intelligence capable of surpassing the capabilities of a single LLM, demonstrate significant performance in mathematical computation \cite{lei2024macm}, code generation \cite{islam2024mapcoder}, software development \cite{he2025llm}, and scientific discovery \cite{ghareeb2025robin}.
% Prior works ~\cite{zhuge2024gptswarm,zhang2025g,li2025assemble} have verified that structured communication topology can improve task processing capability of MAS through collaboration.

\begin{figure}
  \centering
  \includegraphics[width=1\linewidth]{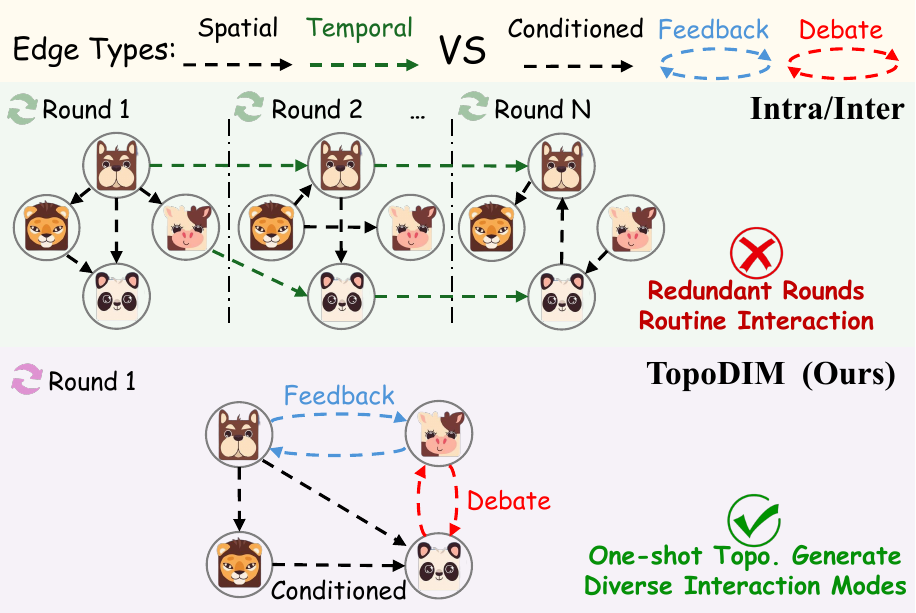}
  \caption{Illustration of hybrid intra/inter-round dialogue method versus \textsc{TopoDIM}. \textsc{TopoDIM} models complex interactions via one-shot topology generation, efficiently cutting potential token overhead.}
  \label{fig:moti}
\end{figure}
Research on MAS with structured communication topologies has predominantly relied on dialogue-based interaction paradigms, as illustrated in Figure~\ref{fig:moti}. In earlier frameworks~\cite{wu2023autogen,du2023improving,zheng2023progressive,liu2024dynamic}, agent connections are homogeneous and restricted to a single interaction mode, with information flow governed by static communication graphs lacking dynamic adaptation. While recent studies~\cite{zhuge2024gptswarm,zhang2025g} have achieved performance gains by hybridizing intra- and inter-round dialogues within spatio-temporal graphs, this multi-round design inherently imposes additional computational and token costs. Although mitigation strategies such as edge pruning~\cite{zhang2024cut}, dynamic agent selection~\cite{wang2025anymac,wang2025agentdropout,li2025assemble}, and proxy-based optimization~\cite{jiang2025dynamic} attempt to balance efficiency and performance, structural redundancy remains unavoidable. Specifically, each inter-round interaction triggers a subsequent intra-round dialogue phase, recursively compounding communication overhead~\cite{chen2025optima,zeng2025s}.

Single-round dialogue schemes offer the advantage of reduced communication overhead; however, maintaining high task performance within such a constrained framework remains a challenge~\cite{hu2026context}. Drawing on recent insights that evaluation and debate mechanisms can improve problem-solving in multi-agent systems \cite{xue2025comas,zhou2025multi}, we posit that explicitly modeling diverse interactions through a heterogeneous communication graph yields a balance between efficiency and performance. Thus, we propose \textsc{TopoDIM} (One-shot \textsc{\textbf{Topo}}logy Generation with \textbf{D}iverse \textbf{I}nteraction \textbf{M}odes). Specifically, \textsc{TopoDIM} is characterized by the following key features:

\ding{182} \textbf{Diverse Interaction Modes}.
\textsc{TopoDIM} leverages three efficient collaborative argumentation primitives \cite{scardamalia2006knowledge}, enabling agents to engage in complementary forms of information exchange beyond homogeneous message passing. \ding{183} \textbf{One-shot Heterogeneous Topology Generation.}
\textsc{TopoDIM} employs a heterogeneous graph encoder to capture agent- and task-specific contexts, coupled with an autoregressive decoder that generates a communication topology with multiple interaction modes in a single inference step, eliminating iterative dialogue rounds. \ding{184} \textbf{Decentralized Architecture.}
To facilitate adaptive decision-making and mitigate privacy risks~\cite{yang2025agentnetdecentralizedevolutionarycoordination}, \textsc{TopoDIM} distills the global topology optimizer into lightweight local networks deployed on individual agents, enabling autonomous and fully decentralized topology construction.

Compared to existing approaches, \textsc{TopoDIM} substantially reducing computational overhead while promoting more diverse collaboration patterns, ultimately leading to improved task-solving performance in multi-agent systems. Our main contributions are summarized as follows:

\begin{itemize}[leftmargin=*, itemsep=-0.3em]
  \item[\ding{182}] \textbf{\textit{Observation.}}
    We suggest that replacing iterative topology generation in both intra- and inter-round dialogue processes with a one-shot heterogeneous topology formulation significantly improves communication efficiency without performance trade-offs.

  \item[\ding{183}] \textbf{\textit{Framework.}}
    We propose \textsc{TopoDIM}, a decentralized framework that integrates a heterogeneous graph encoder with an autoregressive decoder to generate multi-relational communication topologies in one shot, enabling autonomous agent-level decision-making.

  \item[\ding{184}] \textbf{\textit{Evaluation.}}
    Extensive experiments demonstrate that \textsc{TopoDIM} consistently enhances communication efficiency, task performance, and structural robustness across both homogeneous and heterogeneous multi-agent settings, surpassing strong task-adaptive cooperation baselines.
\end{itemize}
\section{Methodology Overview}
\begin{figure*}[t] % 注意这里是 figure*，[t] 表示置顶
  \centering
  % --- 第1张图 ---
  \includegraphics[width=\linewidth]{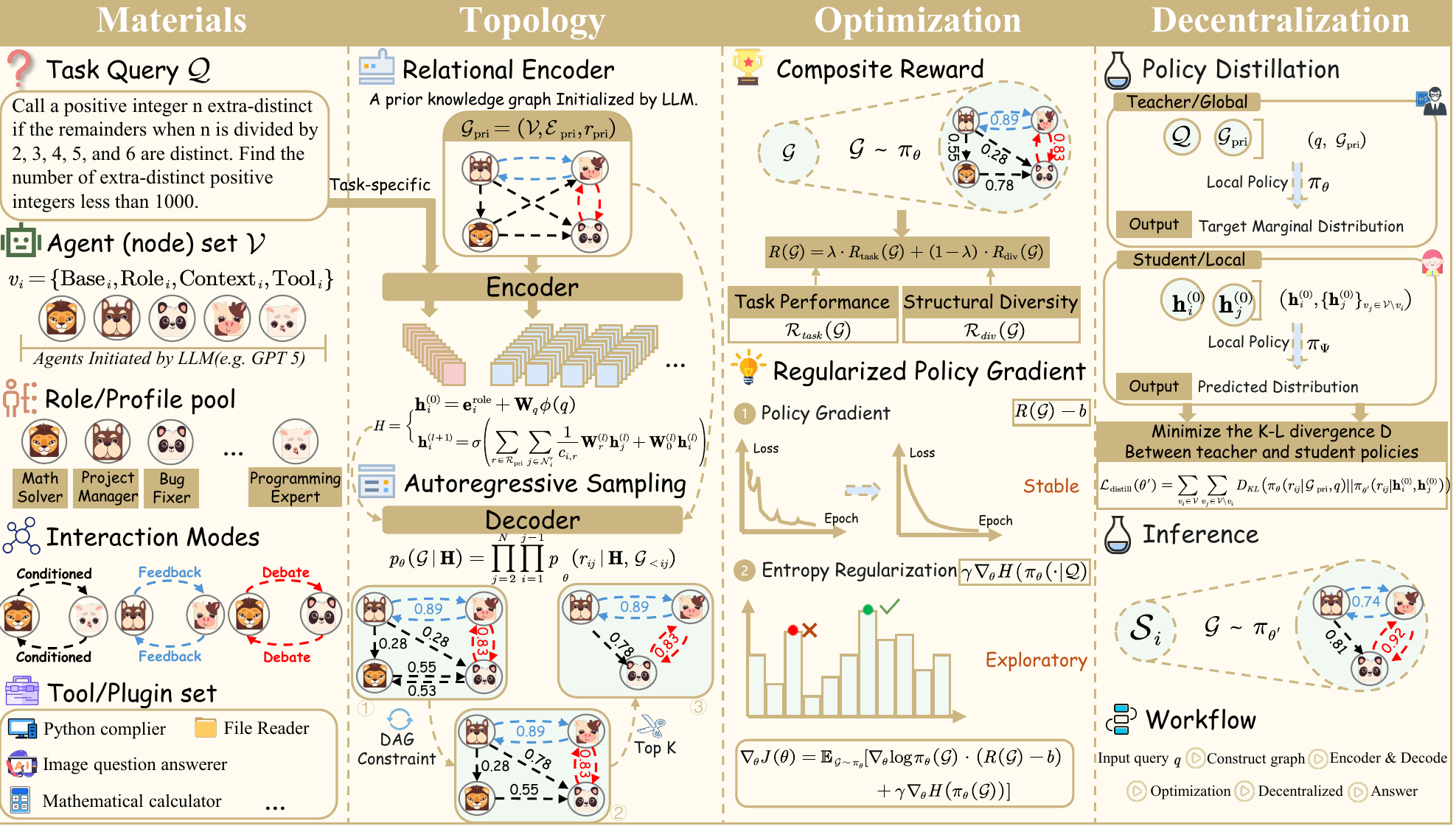}
  \caption{The framework of \textsc{TopoDIM}, comprising 4 components: \ding{182} Materials: defining agents, roles, interaction modes, and plugins; \ding{183} Topology: illustrating the heterogeneous topology design; \ding{184} Optimization: detailing topology optimization strategies, and \ding{185} Decentralization: describing the decentralized agent decision-making.}
  \label{fig:framework}
  % \vspace{-1em}
\end{figure*}

\textsc{TopoDIM} dynamically orchestrates conditioned, feedback, and dialectical behaviors within a single execution to leverage the complementary strengths of LLM agents and heterogeneous interactions. The schematic overview of \textsc{TopoDIM} is presented in Figure~\ref{fig:framework}.
% In this section, we define the symbolic structure of the multi-behavior multi-agent system, introduce the multi-agent communication network pipeline, and discuss the decentralized architecture paradigm.

\subsection{Communication Topology and Pipeline}

% \lm{Explain what are the options for these attributes. For example $\text{Role}_i \in \{\text{math solver}, \cdots\}$, make a reference to Figure 3.}
% \lm{Does two agents can have multiple relations? in this notation it seems yes. Otherwise just define the graph as $G=(V,E,T)$, where $T:E\to R$ is mapping from edges to types}
% \lm{looks this section is not all about communication pipeline? it also talks about optimization. Arrange the content properly.}
\noindent \textbf{Communication topology.} We model LLM-based MAS as a directed heterogeneous graph \(\mathcal{G} = (\mathcal{V}, \mathcal{E}, r)\). The node set \(\mathcal{V} = \{v_1, \dots, v_N\}\) represents the collection of agents. As illustrated in Figure ~\ref{fig:framework}, each agent \(v_i\) is instantiated by a base language model \(\mathrm{LM}_i\), a role \(\mathrm{Role}_i\) (e.g., math-solver), a context \(\mathrm{Context}_i\) which includes the knowledge and dialogues history, and a set of external tools \(\mathrm{Tool}_i\) (e.g., file-reader).
The edge set $\mathcal{E}$ denotes agent interactions, where $r: \mathcal{E} \to \mathcal{R}$ serves as an edge-type mapping function. Specifically, a directed edge $e_{ij}\in\mathcal{E}$  signifies an interaction from $v_i$ to $v_j$ governed by a protocol $r_{ij}:=r(e_{ij})\in\mathcal{R}$. The definition of $\mathcal{R}$ will be discussed in Section~\ref{sec:M1}.

% \lm{Under this notation, the upper-script in $e_{ij}^{(r)}$ is not defined. Also if you use math catholic for sets, here $\mathcal{T}$ is not a set but a function. consider using \(\mathcal{G} = (\mathcal{V}, \mathcal{E},\tau)\). So it should be $\tau (e_{ij}) \in \mathcal{R}$. If you want to lighten the notations, you can write $\tau_{ij}$ }

\noindent \textbf{Communication pipeline.} Given a query \(q\), a scheduling function \(\psi\) maps \(\mathcal{G}\) to an execution sequence \(\psi \rightarrow\sigma = \langle v_{\sigma(1)}, \dots, v_{\sigma(N)} \rangle\). When activated, each agent \(v_i\) processes a prompt \(\mathcal{P}\) to produce a solution \(\mathcal{S}_i = v_i(\mathcal{P})\). The final output \(a\) is obtained by aggregating all agents' solutions: \(a \leftarrow \mathrm{Aggregate}(\{\mathcal{S}_i\}_{i=1}^N)\).
As a comparision, hybrid intra/inter-round dialogue paradigms framework rely on a $T$-round iterative simulation yielding sequential outputs $\{a^{(t)}\}_{t=1}^{T}$, which imposes a potential overhead on execution efficiency.
\subsection{Optimization and Decentralized Design}
% \lm{edge  type missing}
\noindent \textbf{Centralized optimization.} We employ reinforcement learning to optimize the interaction graph. Specifically, we formulate this task as learning a stochastic policy $\pi_{\theta}$ that, given a task query $q$, seeks to construct an optimal interaction graph $\mathcal{G^*}= (\mathcal{V}^{*},\mathcal{E}^{*},r^{*})$. The objective of policy $\pi_{\theta}$ is to maximize the expected reward over the distribution of possible graphs, guided by a task-related reward function $R(\cdot)$ derived from the quality of the final answer:
\begin{equation}
  \begin{split}
    \mathcal{G}^*= \underset{\mathcal{G}}{\operatorname*{argmax}}
    \begin{bmatrix}\mathbb{E}_{\mathcal{G}\sim\pi_\theta(\cdot|q)}[R(\mathcal{G})]
    \end{bmatrix}.
  \end{split}
\end{equation}

% In contrast to prior work\cite{chen2024more,jiang2023llm,qian2024scaling} utilizing centralized architectures, which predefine the global communication topology, d
\noindent \textbf{Decentralized design.} To support adaptive decision-making while addressing potential privacy risks, we adopt a decentralized architecture where each agent $v_i \in \mathcal{V}$ maintains an independent local policy network $\pi^{(i)}_{\theta^\prime}$ to infer its connectivity with other nodes. Conditioned on the task embedding $\mathbf{z}$ and the local states $\mathbf{h}_i$ and $\mathbf{h}_j$, the communication link between agent $v_i$ and $v_j$ is formulated as:
\begin{equation}
  p_{ij}^{(r)}=\pi_{\theta^\prime}^{(i)}(r_{ij}|\mathbf{h}_i,\mathbf{h}_j,\mathbf{z}).
\end{equation}

\section{Architecture Details}
% Drawing inspiration from recent advances in multi-agent collaboration \cite{xue2025comas,choi2025debate},
% \lm{Explain the three ways of interaction in more detail, also why you design this way.}

\subsection{Diverse Interaction Modes}\label{sec:M1}
\begin{figure}
  \centering
  \includegraphics[width=1\linewidth]{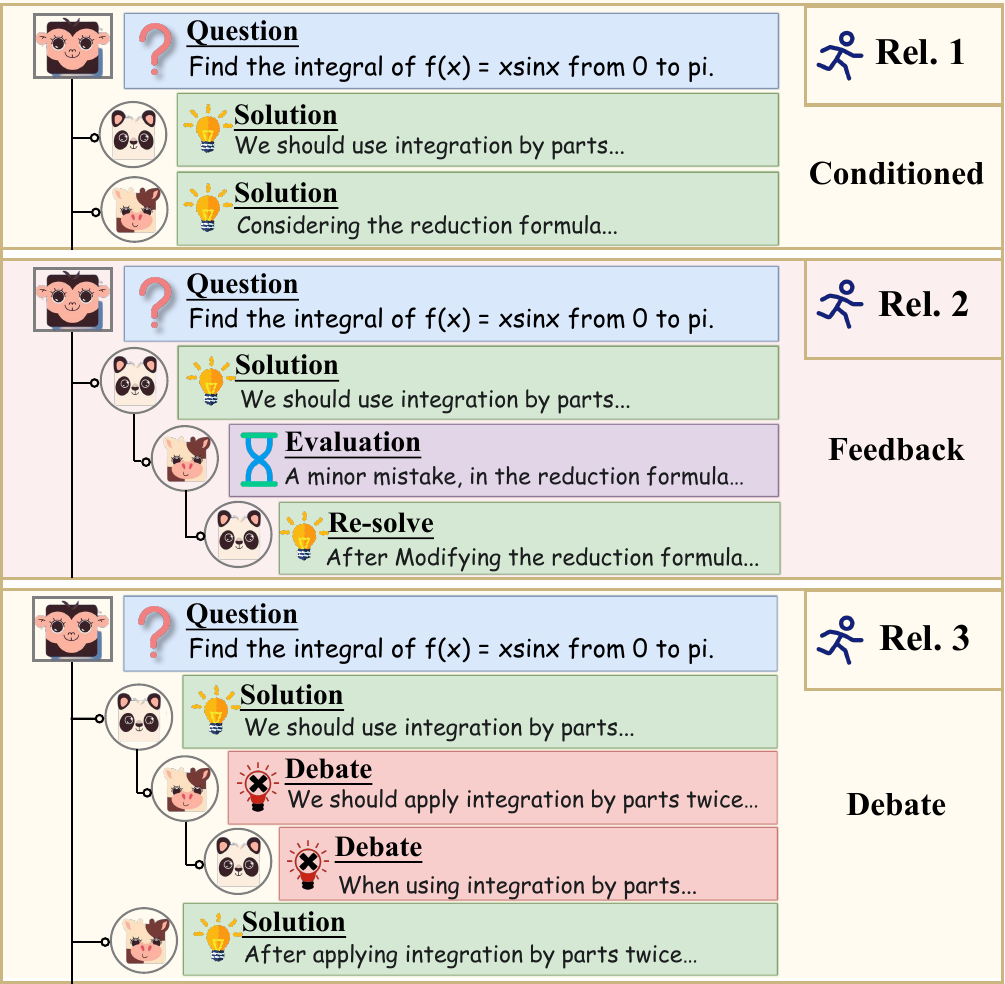}
  % \caption[Interaction modes of \textsc{TopoDIM}]{Interaction modes of \textsc{TopoDIM}. Information flow for $\{\mathrm{conditioned, debate}\}$ proceeds from \raisebox{-0.3em}{\includegraphics[height=1.3em]{figure/icon1.pdf}} to \raisebox{-0.3em}{\includegraphics[height=1.3em]{figure/icon2.pdf}}, while $\mathrm{feedback}$ starts and ends at \raisebox{-0.3em}{\includegraphics[height=1.3em]{figure/icon1.pdf}}.}
  \caption{Interaction modes of \textsc{TopoDIM}. \textsc{TopoDIM} selects three effective interaction modes aiming to optimize the execution sequence among the agents.}
  \label{fig:interaction}
\end{figure}

Recent studies indicate that diverse collaborative mechanisms, such as evaluation and debate, can significantly bolster the complex problem-solving capabilities of MAS \cite{xue2025comas,zhou2025multi}. Motivated by these findings, we formalize the interaction space $\mathcal{R}$ using three distinct edge types, as shown in Figure~\ref{fig:interaction}.
Each edge type is instantiated via specific prompts to govern its interaction logic.

\noindent \ding{182} \textbf{Conditioned edges.} The classical edge type where agent $v_j$ handles the query $q$ conditioned with the outputs of agent $v_i$.

\noindent \ding{183} \textbf{Feedback edges.} Encapsulate a supervisory mechanism where agent $v_j$ critiques or validates the intermediate outputs generated by agent $v_i$ and $v_i$ re-handle the query $q$ referring to the feedback of $v_j$, simulating an evaluation and reflection process.

\noindent \ding{184} \textbf{Debate edges.} Model a debate process wherein agent $v_j$ challenges $v_i$'s proposition for two rounds, ultimately $v_j$ proceed the query $q$ with the context of the debate.

%\noindent We note that the distribution of edge types $\mathcal{R}_{\text{constr}}= {\{\mathrm{conditioned,debate}\}}$ affects acyclic structural constraints, and how to force the acyclic constraint during graph inference will be discussed in Section~\ref{sec:M2}. Meanwhile, $\mathcal{R}_{\text{free}} = \{\mathrm{feedback}\}$ is exempt from this restriction.
%The reason for design of $\mathcal{R}_{\text{free}}$ and $\mathcal{R}_{\text{constr}}$ is provided in Appendix \ref{sec:edge-supplement}.

% The interaction graph is traversed in a breadth-first manner. Starting from the root node, each agent interacts with its neighbors following a predefined order, sorted by ascending node indices. After an interaction, neighbors connected via feedback edges are removed from the graph, whereas neighbors connected through the other two interaction types are retained and scheduled for future visits. As a result, feedback edges do not introduce circular dependencies during traversal. A detailed discussion of how circular dependencies are avoided is provided in Section~3.2.

Given a query $q$, \textsc{TopoDIM} aims to learn a policy $\pi_{\theta}$ that determines the optimal relation $r^*_{ij} \in \mathcal{R}^\prime:= \mathcal{R}\cup \{\emptyset\}$ between the agent pair $(v_i, v_j)$, where $\emptyset$ signifies the non-existence of an edge.
\begin{equation}
  r^*_{ij} = \underset{r \in \mathcal{R}^\prime}{\arg\max} \ \pi_{\theta}\left(r \vert v_i, v_j, q \right).
\end{equation}

\subsection{Heterogeneous Interaction Topology}
\label{sec:M2}
% \lm{looks vert is more compact than mid}
We formulate the generation of heterogeneous topologies as a conditional autoregressive decision process. The learnable policy $\pi_\theta$ defines a distribution $p_{\theta}(\mathcal{G} \vert q)$ over feasible communication graphs.

\noindent \textbf{Semantics-aware relational encoder.}
To establish a robust latent representation for each agent, we employ a relational graph convolutional network \cite{schlichtkrull2018modeling,wang2024llm} over a prior knowledge graph $\mathcal{G}_\text{pri} = (\mathcal{V},\mathcal{E}_{\mathrm{pri}},r_{\mathrm{pri}})$, which is initialized by an advanced LLM (e.g., GPT-5). The node update rule at layer $l+1$ is defined as $\mathbf{h}^{\ell+1}_i = \sigma (\hat{\mathbf{h}}^{\ell}_i)$, where
\begin{equation}
  \hat{\mathbf{h}}^{\ell}_i = \sum_{r \in \mathcal{R}_{\mathrm{pri}}} \sum_{j \in \mathcal{N}_i^r} \frac{1}{c_{i,r}} \mathbf{W}_r^{(l)} \mathbf{h}_j^{(l)} + \mathbf{W}_0^{(l)} \mathbf{h}_i^{(l)},
\end{equation}
$\mathcal{N}_{i}^{r}$ denotes the set of neighbors of node $v_i$ under relation $r$, and $c_{i,r}=|\mathcal{N}_i^r|$. The initial node embeddings combine role and task information:
\begin{equation}
  \mathbf{h}_i^{(0)}=\mathbf{e}_i^\text{role}+\mathbf{W}_q\phi(q),
\end{equation}
where $\mathbf{e}_{i}^{\mathrm{role}}$ is a learnable role embedding and $\phi(q)$ is the feature representation of the task query $q \in Q$ obtained from a pre-trained sentence encoder. The final set of node embeddings from the encoder, $\{\mathbf{h}_1, \dots, \mathbf{h}_N\}$, constitutes the conditional representation $\mathbf{H}$.

\noindent \textbf{Autoregressive edge sampling decoder.} Given the conditional representation $\mathbf{H}$, a decoder generates topology $\mathcal{G}$ sequentially. With a prespecified ordering of nodes, the joint probability of observing a graph $\mathcal{G}$ reads:
\begin{equation}
  \begin{split}
    &p_\theta(\mathcal{G} \mid \mathbf{H}) = \prod_{j=2}^N \prod_{i=1}^{j-1} p_{\theta}(r_{ij} \mid \mathbf{H}, \mathcal{G}_{<ij}) \\
    &= \prod_{j=2}^N \prod_{i=1}^{j-1} \frac{\exp\left(\mathbf{w}_{r_{ij}}^\top [\mathbf{h}_i \| \mathbf{h}_j] / \tau\right)}{\sum_{r \in \mathcal{R}^\prime} \exp\left(\mathbf{w}_r^\top [\mathbf{h}_i \| \mathbf{h}_j] / \tau\right)},
  \end{split}
\end{equation}
where $\|$ denotes concatenation, and $\tau$ is the temperature.
We sample the relation type $r_{ij}$ using inverse transform sampling by generating $u \sim \text{Uniform}(0,1)$ and setting $F(r) = \sum_{r\in\mathcal{R}^\prime} p_\theta(r)$, where $r_{ij} = \min\{r : F(r) \geq u\}$ represents the cumulative probability.

\noindent \textbf{Structural constraints.}
To ensure logical consistency and prevent circular dependencies, we enforce an acyclic constraint via a dynamic mask $\mathbf{M}$. Specifically, if the presence of edge $(v_i,v_j)$ violates the acyclic constraint in any existing path from $v_i$ to $v_j$,  the probability of this node pair is masked as:
\begin{equation}
  p(r_{ij} \mid \cdot) \leftarrow p(r_{ij} \mid \cdot) \odot \mathbf{M}_{ij},
\end{equation}
where $\mathbf{M}_{ij} = \mathbb{I}(v_i \not\rightsquigarrow v_j)$ indicates if $v_i$ and $v_j$ are not path-connected. Detailed discussions on structural constraint is shown in Appendix~\ref{sec:edge-supplement}.

\noindent \textbf{Adaptive sparsification.} Inspired by connectivity sparsification methods \cite{zhang2024cut,wang2025agentdropout} that enhance task performance through structural reduction, we employ an adaptive pruning mechanism to retain only salient interactions. Specifically, we identify the subset of edges $\mathcal{E}_{\mathrm{final}}$ based on confidence scores, subject to a sparsity budget ratio $\alpha$:
\begin{equation}
  \begin{split}
    &\mathcal{E}_{\mathrm{final}}=\mathrm{TopK}\left(\{p(r_{ij})\}_{\forall i,j},1-\alpha\right).
  \end{split}
\end{equation}
To identify the most relevant contributors for each task, \textsc{TopoDIM} filters out inactive nodes that lack connectivity within the selected edge subset. The final node set $\mathcal{V}_{\mathrm{final}}$ is derived by retaining only those vertices involved in at least one valid interaction:
\begin{equation}
  % \mathcal{V}_{\mathrm{final}} = \left\{ v \in \mathcal{V} \;\middle|\;
  % \begin{aligned}
  % &\exists u \in \mathcal{V}, r \in \mathcal{R} : \\
  % &(v, u, r) \in \mathcal{E}_{\mathrm{final}} \; \lor \\
  % &(u, v, r) \in \mathcal{E}_{\mathrm{final}}
  % \end{aligned}
  % \right\}.
  \mathcal{V}_{\mathrm{final}}=\{v\mid\exists(h,t,r)\in\mathcal{E}_{\mathrm{final}},v\in\{h,t\}\}.
\end{equation}

\noindent \textbf{Execution order.} Upon establishing the final set of agents $\mathcal{V}_{\mathrm{final}}$ and edges $\mathcal{E}_{\mathrm{final}}$, the interaction process follows a breadth-first manner. Starting from the root node (characterized by zero in-degree), each agent interacts with its neighbors in a sequence determined by ascending node indices.

% \subsection{Stabilized and Diversity-Aware Optimization Strategy}
\subsection{Diversity-Aware Optimization Strategy}
We optimize a graph generation policy $\pi_{\theta}$ via reinforcement learning, aiming to maximize a composite objective that balances task success with structural diversity. For each generated graph $\mathcal{G} \sim \pi_{\theta}$, the agent ensembles execute the collaborative task to yield a binary performance metric $R_{\mathrm{task}}(\mathcal{G})\in\{0,1\}$, where 1 indicates a successful solution and 0 indicates failure.
To mitigate mode collapse and prevent the policy from converging to a narrow set of interaction patterns, we introduce a structural balance reward~\cite{ma2026tspo,liu2026bapo}. This term is formulated as the Shannon entropy of the empirical edge-type distribution $p(r) = |\mathcal{E}_r|/|\mathcal{E}|$, where $\mathcal{E}_r$ denotes the set of edges with relation type $r$:
\begin{equation}
  R_{\text{div}}(\mathcal{G}) = - \sum_{r\in \mathcal{R}} p(r) \log p(r).
\end{equation}
The holistic reward signal $R(\mathcal{G})$ is a weighted interpolation of task performance and structural entropy:
\begin{equation}
  R(\mathcal{G}) = \lambda \cdot R_{\text{task}}(\mathcal{G}) + (1-\lambda) \cdot R_{\text{div}}(\mathcal{G}),
\end{equation}
where $\lambda \in [0,1]$ is a hyperparameter governing the trade-off between performance exploitation and diversity exploration.

Following the policy gradient theorem \cite{sutton1999policy}, we maximize the expected objective $J(\theta) = \mathbb{E}_{\mathcal{G}\sim\pi_\theta}[R(\mathcal{G})]$. To stabilize training and reduce variance, we employ a moving average baseline $b$ \cite{williams1992simple,chen2024decentralized}.
The overall loss function also includes an entropy regularization term for the policy's output distribution to encourage exploration:
\begin{equation}
  \begin{split}
    \nabla_\theta J(\theta) = & \mathbb{E}_{\mathcal{G}\sim\pi_\theta} \Big[ \nabla_\theta \log \pi_\theta(\mathcal{G}) \cdot \left( R(\mathcal{G}) - b \right) \\
    &+ \gamma \nabla_\theta H(\pi_\theta(\mathcal{G})) \Big],
  \end{split}
\end{equation}
where $H(\pi_\theta)$ denotes the entropy of the policy distribution and $\gamma$ is the regularization coefficient.

\subsection{Decentralized Architecture Design}
To facilitate scalable and autonomous deployment, each agent $v_i$ employs a light local policy (e.g., MLP) $\pi_{\theta^\prime}^{(i)}$ to predict a relation type with all other node $\sum_{j\neq i}v_j$, taking each ordered pair of representations$( \mathbf{h}_i^{(0)},\mathbf{h}^{(0)}_j)$. We align local policy with the global policy $\pi_{\theta}$ via policy distillation, minimizing the Kullback-Leibler (KL) divergence $D_{KL}$ to capture global structural dependencies:
% \begin{equation}
% \begin{split}
% & \mathcal{L}_{\text{distill}}(\Psi) \\
% &= \sum_{\text{task}} \sum_{i=1}^{|\mathcal{V}|} D \left( \pi_\theta(\cdot | \mathcal{G}_{pri}, q) \parallel \pi_{\Psi}^{(i)}(\cdot | \mathbf{s}_i) \right),
% \end{split}
% \end{equation}
\begin{equation}
  \begin{split}
    \mathcal{L}_{\text{distill}}&(\theta^\prime) =  {\sum_{v_i \in \mathcal{V}}}{\sum_{v_j \in \mathcal{V} \setminus v_i}} D_{KL}\big( \\
    &\pi_\theta (r_{ij}| \mathcal{G}_{\text{pri}},q) || \pi_{\theta^\prime}(r_{ij} | \mathbf{h}^{(0)}_i, \mathbf{h}^{(0)}_j)\big),
  \end{split}
\end{equation}
where $\pi_\theta(\cdot|\cdot)$ is the centralized policy induced marginal distribution. During inference, agents autonomously sample connections from $\pi_{\theta^\prime}^{(i)}$ following the structural constraints, thereby combining centralized training optimality with decentralized execution efficiency.

\begin{table*}[!t]
  \centering
  % \caption{Performance comparison with three types of baselines, including single-agent execution, static multi-agent topologies, and adaptive multi-agent frameworks. The best results are in bold, and the runner-ups are underlined. {All multi-agent methods utilize \textbf{five} \llmname{gpt-4}-based agents.} ``Mul.'', ``Ada.'', and ``Rob.'' indicate whether the method supports a multi-agent setting, whether it is task-adaptive, and whether it is adversarially robust, respectively. \textcolor{darksalmon}{\faTimes}, {\large\textcolor{Dandelion}{{\faCheck}{\small{\kern-0.85em\faTimes}}}} and \textcolor{greenpigment}{\faCheck} signifies no/partial/full support in these aspects.}
  \caption{Performance comparison with three different base LLMs including \texttt{Gemma-3-it:12B}, \texttt{GPT-OSS-120B}, and \texttt{DeepSeek-V3.2}. \textbf{Ada.}, \textbf{DIM.}, and \textbf{Dec.} indicates whether it is task-adaptive, whether it supports diverse interaction modes, and whether it is decentralized architecture respectively. The \textbf{bold} and \underline{underlined} fonts show the best and the second best result. \textcolor{darksalmon}{\faTimes}, {\large\textcolor{Dandelion}{{\faCheck}{\small{\kern-0.85em\faTimes}}}},  and \textcolor{greenpigment}{\faCheck} signifies no/partial/full support in these aspects.}
  \vspace{-0.1em}
  \label{table1}
  \renewcommand\tabcolsep{5.3pt}
  \renewcommand\arraystretch{1.1}

  \resizebox{\linewidth}{!}{
    \begin{tabular}{l|ccc|ccccccc}
      \Xhline{1.2pt}
      \rowcolor{CadetBlue!20}
      {\textbf{Method}} & \textbf{Ada.} & \textbf{DIM.} & \textbf{Dec.}& \textbf{LiveCodeBench} & \textbf{MMLU-Pro} & \textbf{GSM8K} & \textbf{MultiArith} & \textbf{AIME} & \textbf{HumanEval} & {\textbf{Avg.}} \\
      \Xhline{1.2pt}
      \multicolumn{11}{c}{\textit{Base model: Gemma-3-it:12B}} \\
      \hline
      Vanilla & \textcolor{darksalmon}{\faTimes} & \textcolor{darksalmon}{\faTimes} & \textcolor{darksalmon}{\faTimes} & 24.46 & 51.24 & 87.37 & 96.68 & 11.28 & 82.23 & 58.88\\

      \rowcolor{gray!10}CoT & \textcolor{darksalmon}{\faTimes}  & \textcolor{darksalmon}{\faTimes} & \textcolor{darksalmon}{\faTimes} & 24.90\red{0.44} & 51.95\red{0.71} & 87.83\red{0.46} & 97.01\red{0.33} & 12.14\red{0.86} & 82.85\red{0.62} & 59.45\red{0.57}\\

      \hdashline

      LLM-Debate & \textcolor{darksalmon}{\faTimes}  &\textcolor{darksalmon}{\faTimes} & \textcolor{darksalmon}{\faTimes} & 25.47\red{1.01} & 52.66\red{1.42} & 88.46\red{1.09} & 97.62\red{0.94} & 11.88\red{0.60} & 83.85\red{1.62} & 59.99\red{1.11}\\

      \rowcolor{gray!10}GPTSwarm & {\large\textcolor{Dandelion}{{\faCheck}{\small{\kern-0.85em\faTimes}}}}   & \textcolor{darksalmon}{\faTimes} & \textcolor{darksalmon}{\faTimes} & 26.22\red{1.76} & 53.14\red{1.90} & 89.38\red{2.01} & 97.85\red{1.17} & 12.50\red{1.22} & 84.14\red{1.91} & 60.54\red{1.66} \\

      AgentDropout & \textcolor{greenpigment}{\faCheck}  & {\large\textcolor{Dandelion}{{\faCheck}{\small{\kern-0.85em\faTimes}}}} & \textcolor{darksalmon}{\faTimes} & 26.75\red{2.29} & 53.79\red{2.55} & 90.14\red{2.77} & 98.20\red{1.52} & \underline{13.24}\red{1.96} & 84.73\red{2.50} & 61.14\red{2.27}\\

      \rowcolor{gray!10} G-Designer & \textcolor{greenpigment}{\faCheck}   & {\large\textcolor{Dandelion}{{\faCheck}{\small{\kern-0.85em\faTimes}}}} & \textcolor{darksalmon}{\faTimes} & \underline{27.08}\red{2.62} & \underline{54.25}\red{3.01} & \underline{90.27}\red{2.90} & \underline{98.45}\red{1.77} & 12.86\red{1.58} & \underline{85.47}\red{3.24} & \underline{61.40}\red{2.52}\\

      \rowcolor{lightpurple}  \textbf{\textsc{TopoDIM}} & \textcolor{greenpigment}{\faCheck}  & \textcolor{greenpigment}{\faCheck} & \textcolor{greenpigment}{\faCheck} & \textbf{27.38}\red{2.92} & \textbf{55.08}\red{3.84} & \textbf{91.86}\red{4.49} & \textbf{98.85}\red{2.17} & \textbf{13.58}\red{2.30} & \textbf{86.62}\red{4.39} & \textbf{62.23}\red{3.35}\\

      \hline
      \multicolumn{11}{c}{\textit{Base model: GPT-OSS:120B}} \\
      \hline

      Vanilla & \textcolor{darksalmon}{\faTimes} & \textcolor{darksalmon}{\faTimes} & \textcolor{darksalmon}{\faTimes}& 81.58 & 74.49
      & 95.90 & 100 & 74.77 & 90.04 & 86.13\\

      \rowcolor{gray!70} CoT & \textcolor{darksalmon}{\faTimes}  & \textcolor{darksalmon}{\faTimes} & \textcolor{darksalmon}{\faTimes}&  &  &&&&&\\

      \hdashline

      LLM-Debate &  \textcolor{darksalmon}{\faTimes}   &\textcolor{darksalmon}{\faTimes} & \textcolor{darksalmon}{\faTimes} & 82.92\red{1.34} & 75.53\red{1.04} & 95.72\blue{0.18} & 100\red{0.00} & 75.38\red{0.61} & 91.43\red{1.39} & 86.83\red{0.70}\\

      \rowcolor{gray!10}GPTSwarm & {\large\textcolor{Dandelion}{{\faCheck}{\small{\kern-0.85em\faTimes}}}}   & \textcolor{darksalmon}{\faTimes} & \textcolor{darksalmon}{\faTimes} & 83.59\red{2.01} & 76.88\red{2.39} & 96.50\red{0.60} & 100\red{0.00} & 77.41\red{2.64} & 92.30\red{2.26} & 87.78\red{1.65}\\

      AgentDropout & \textcolor{greenpigment}{\faCheck}  & {\large\textcolor{Dandelion}{{\faCheck}{\small{\kern-0.85em\faTimes}}}} & \textcolor{darksalmon}{\faTimes}& 85.06\red{3.48} & 78.62\red{4.13} & 96.86\red{0.96} & 100\red{0.00} & 78.13\red{3.36} & 92.88\red{2.84} & 88.59\red{2.46}\\

      \rowcolor{gray!10}G-Designer & \textcolor{greenpigment}{\faCheck}   & {\large\textcolor{Dandelion}{{\faCheck}{\small{\kern-0.85em\faTimes}}}} & \textcolor{darksalmon}{\faTimes} & \underline{85.63}\red{4.05} &\underline{79.08}\red{4.59} & \underline{97.08}\red{1.18} & \underline{100}\red{0.00} & \underline{78.62}\red{3.85} & \underline{93.47}\red{3.43} & \underline{88.98}\red{2.85}\\

      \rowcolor{lightpurple}  \textbf{\textsc{TopoDIM}} & \textcolor{greenpigment}{\faCheck}  & \textcolor{greenpigment}{\faCheck} & \textcolor{greenpigment}{\faCheck} & \textbf{87.28\red{5.70}} & \textbf{80.11}\red{5.62} & \textbf{{98.34}\red{2.44}} & \textbf{100}\red{0.00} & \textbf{80.34}\red{5.57} & \textbf{95.83}\red{5.79} & \textbf{90.32}\red{4.19}\\

      \hline
      \multicolumn{11}{c}{\textit{Base model: DeepSeek-V3.2-251201:671B}} \\
      \hline

      Vanilla & \textcolor{darksalmon}{\faTimes} & \textcolor{darksalmon}{\faTimes} & \textcolor{darksalmon}{\faTimes}& 78.35  & 78.68 & 96.31 & 100 & 64.37 & 89.46 & 83.53\\

      \rowcolor{gray!10}CoT & \textcolor{darksalmon}{\faTimes}  & \textcolor{darksalmon}{\faTimes} & \textcolor{darksalmon}{\faTimes}& 78.84\red{0.49} & 78.92\red{0.24} & 96.58\red{0.27} & 100\red{0.00} & 64.08\blue{0.29} & 89.84\red{0.38} & 84.71\red{0.18} \\

      \hdashline

      LLM-Debate &  \textcolor{darksalmon}{\faTimes}   &\textcolor{darksalmon}{\faTimes} & \textcolor{darksalmon}{\faTimes} & 79.66\red{1.31} & 80.14\red{1.46} & 96.82\red{0.51} & 100\red{0.00} & 64.84\red{0.47} & 90.67\red{1.21} & 85.36\red{0.83}\\

      \rowcolor{gray!10}GPTSwarm & {\large\textcolor{Dandelion}{{\faCheck}{\small{\kern-0.85em\faTimes}}}}   & \textcolor{darksalmon}{\faTimes} & \textcolor{darksalmon}{\faTimes}& 80.40\red{2.05} & 81.75\red{3.07} & 97.14\red{0.83} & 100\red{0.00} & 67.20\red{2.83} & 92.32\red{2.86} & 86.47\red{1.94} \\

      AgentDropout & \textcolor{greenpigment}{\faCheck}  & {\large\textcolor{Dandelion}{{\faCheck}{\small{\kern-0.85em\faTimes}}}}& \textcolor{darksalmon}{\faTimes}& 81.19\red{2.84} & 82.37\red{3.69} & \underline{97.76}\red{1.45} & 100\red{0.00} & 68.17\red{3.80} & 93.15\red{3.69} & 87.11\red{2.58}\\

      \rowcolor{gray!10}G-Designer & \textcolor{greenpigment}{\faCheck}   & {\large\textcolor{Dandelion}{{\faCheck}{\small{\kern-0.85em\faTimes}}}} & \textcolor{darksalmon}{\faTimes} & \underline{81.63}\red{3.28} & \underline{82.83}\red{4.15} & 97.48\red{1.17} & \underline{100}\red{0.00} & \underline{68.62}\red{4.25} & \underline{93.54}\red{4.08} & \underline{87.35}\red{2.82}\\

      \rowcolor{lightpurple}  \textbf{\textsc{TopoDIM}} & \textcolor{greenpigment}{\faCheck}  & \textcolor{greenpigment}{\faCheck} & \textcolor{greenpigment}{\faCheck}& \textbf{83.26}\red{4.91} & \textbf{84.80}\red{6.12} & \textbf{98.52}\red{2.21} & \textbf{100}\red{0.00} & \textbf{69.93}\red{5.56} & \textbf{94.86}\red{5.20} & \textbf{88.53}\red{4.03}\\

      \Xhline{1.2pt}
    \end{tabular}
  }
\end{table*}

\section{Experiment}
% To validate the effectiveness of our proposed \textsc{TopoDIM} framework, we try to answer the following questions:

% \noindent \textbf{RQ1:} How effective does \textsc{TopoDIM} perform?

% \noindent \textbf{RQ2:} How is the adaptability of heterogeneous agents with \textsc{TopoDIM}?

% \noindent \textbf{RQ3:} How is the cost-efficiency of \textsc{TopoDIM}?

% \noindent \textbf{RQ4:} How do the key components of \textsc{TopoDIM} affect its performance?

% \noindent \textbf{RQ5:} How does \textsc{TopoDIM} solve task explainability and efficiently?
% \texttt{Qwen3-32B-Instruct}\cite{yang2025qwen3},
\subsection{Experimental Setups}
\noindent \textbf{Models and benchmarks.} To evaluate the general adaptability of \textsc{TopoDIM} to different LLMs, we employ a diverse set of models including \texttt{Gemma-3-it:12B} \cite{team2025gemma}, \texttt{GPT-OSS-20B} \cite{agarwal2025gpt}, \texttt{GPT-OSS-120B} \cite{agarwal2025gpt}, and \texttt{DeepSeek-V3.2-251201:671B} \cite{liu2025deepseek}.
We test the performance of \textsc{TopoDIM} on three types of reasoning datasets:
\ding{182} General Reasoning: MMLU-Pro \cite{wang2024mmlu};
\ding{183} Mathematics: MultiArith \cite{roy2016solving}, GSM8K \cite{cobbe2021training} , AIME$_{(2023-2025)}$; \ding{184} Coding: HumanEval \cite{hendrycks2020measuring};
LiveCodeBench$_{(202305-202403)} $\cite{jain2024livecodebench}.

\noindent \textbf{Baselines.} We evaluate \textsc{TopoDIM} by comparing its task-solving performance against 6 typical methods, including \ding{182} Single-agent prompting methods: Vanilla, Chain-of-Thought \cite{wei2022chain} which augment few-shot exemplars with intermediate reasoning steps.
\ding{183} Multi-agent topology methods:
LLM-Debate \cite{du2023improving},
GPTSwarm \cite{zhuge2024gptswarm},
G-Designer \cite{zhang2025g},
AgentDropout \cite{wang2025agentdropout}.

\noindent \textbf{Implementations.}  In our experiments, \texttt{Gemma-3} and \texttt{GPT-OSS} are deployed locally on an NVIDIA RTX 4090 via Ollama~\cite{ollama}, while \texttt{DeepSeek-V3.2} is accessed through its official APIs. We implement a temperature annealing schedule decaying from 2.0 to 0.5 and maintain a fixed learning rate of 0.01 for both centralized and decentralized training. Specific parameters include Top-K sparsity budget ratio $\alpha \in \{0.3, 0.5, 0.7\}$, $\lambda \in \{0.3, 0.5, 0.8\}$, and $\eta \in \{0.01,0.03,0.05\}$. The feature extraction function $\phi(\cdot)$ is instantiated with \texttt{all-MiniLM-L6-v2} \cite{wang2020minilm}, yielding $384$-dimensional embeddings.
Each experimental run is initialized with 5 agents. For all benchmarks, the query budgets for optimization ($M$) and and decentralized knowledge distillation ($M'$) are selected from $\{40, 80\}$. Furthermore, we employ \texttt{GPT-5}\cite{openai2025gpt5} to generate prompt descriptions for three distinct interaction types, and the local policy is parameterized by an MLP. All remaining configurations and agent settings follow the protocols established in \cite{zhang2025g}.

% This illustrates that the \textsc{TopoDIM} framework can fully unleash the inherent potential of high-performance LLMs as intelligent agents, achieving synergistic growth between system performance and agent capabilities.

\subsection{Main Results}
\noindent \textbf{Task performance.} Table \ref{table1} demonstrates that \textsc{TopoDIM} outperforms various baselines across multiple datasets, including single models and state-of-the-art multi-agent systems. Specifically, with \texttt{Gemma-3-it:12B} and \texttt{DeepSeek-V3.2-251201}, \textsc{TopoDIM} yields average performance gains of 1.35\% $\uparrow$ and 1.38\% $\uparrow$ over the strongest baselines, respectively. Notably, \textsc{TopoDIM} also enhances the capabilities of the reasoning model (\texttt{GPT-OSS:120B}) on complex tasks, resulting in a 1.50\% $\uparrow$ increase in average predictive accuracy.

\begin{figure}
  \centering
  \includegraphics[width=1\linewidth]{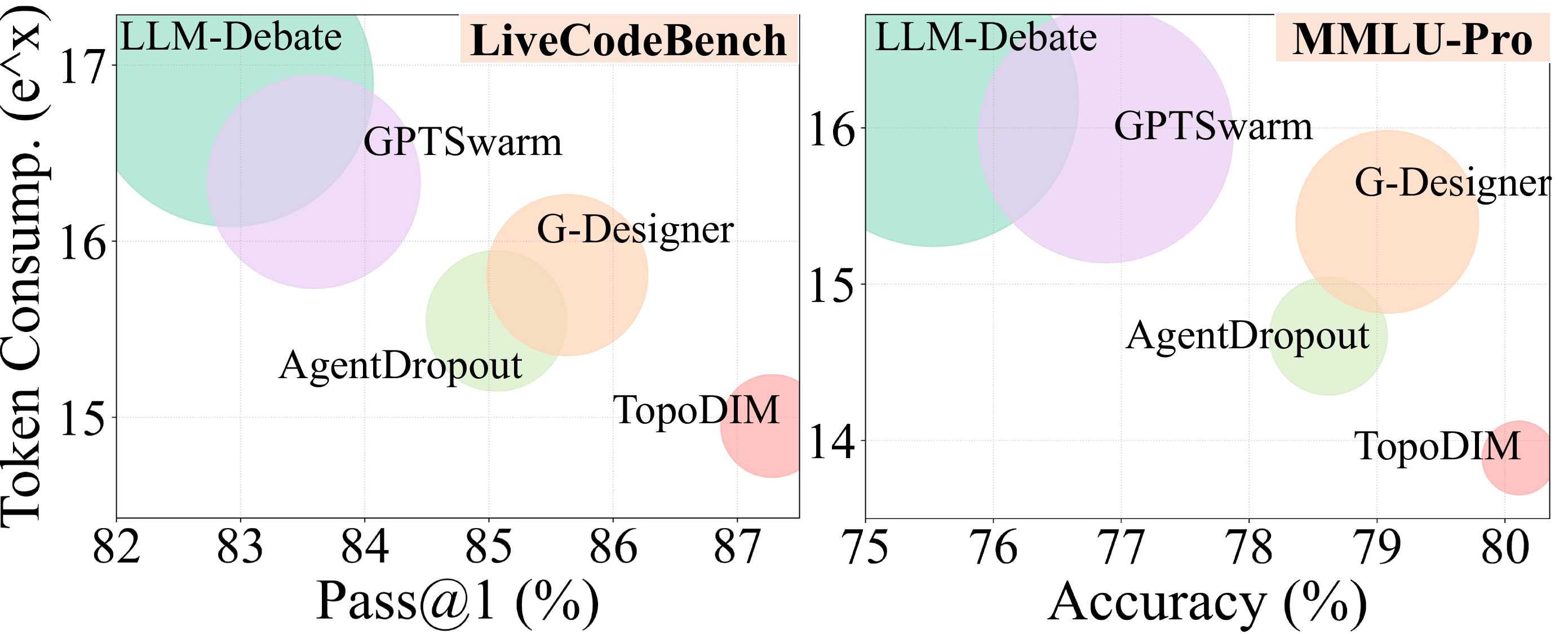}
  \caption{Performance-cost trade-off between \textsc{TopoDIM} and state-of-the-art methods. The bubble size is proportional to token consumption.}
  \label{fig:cost}
  % \vspace{-0.5em}
  % \setlength{\belowcaptionskip}{0pt}
\end{figure}

% \noindent \textbf{Task performance.} Results in Table \ref{table1} indicate that \textsc{TopoDIM} outperforms other baselines on multiple datasets, including single models and SOTA multi-agent structured topology systems. Specifically, when using \texttt{Gemma-3-it:12B} and \texttt{DeepSeek-V3.2-251201}, average performance is improved by 1.35\% $\uparrow$ and 1.38\% $\uparrow$ compared to the most advanced methods. It is noteworthy that \textsc{TopoDIM} improves the performance of reasoning model (\texttt{GPT-OSS:120B}) on complex tasks, improving average predictive performance by 1.50\% $\uparrow$.
% \textbf{Ctok.} and \textbf{Ptok.} indicate prompt and complete token consumption, respectively}
\noindent \textbf{Computation efficiency.} We compare the token consumption of \textsc{TopoDIM} against other baselines on MMLU-Pro and LiveCodeBench using \texttt{GPT-OSS:120B}, with the results illustrated in Figure~\ref{fig:cost}. \textsc{TopoDIM} significantly reduces token expenditure, saving a total of 1.42M and 2.44M tokens on MMLU-Pro and LiveCodeBench, respectively, while maintaining competitive performance. A comprehensive efficiency analysis is provided in Appendix~\ref{sec:cost-ef}.
% \noindent \textbf{Cost efficiency.} We calculate the token consumption of \textsc{TopoDIM} and other baselines on MMLU-Pro and LiveCodeBench using \texttt{GPT-OSS:120B}, which is shown in Figure~\ref{fig:cost}. \textsc{TopoDIM} cuts down the token expenditure with a total token savings amount of 1.46M and 2.4M on MMLU-Pro and LiveCodeBench, simultaneously achieving a competitive performance. More detailed efficiency demonstration is illustrated in Appendix~\ref{sec:cost-ef}.

% saving maximum prompt and completion token by 57.82\% and 22.05\% compared to the most efficient framework,

\begin{figure*} % 注意这里是 figure*，[t] 表示置顶
  \centering
  \includegraphics[width=\linewidth]{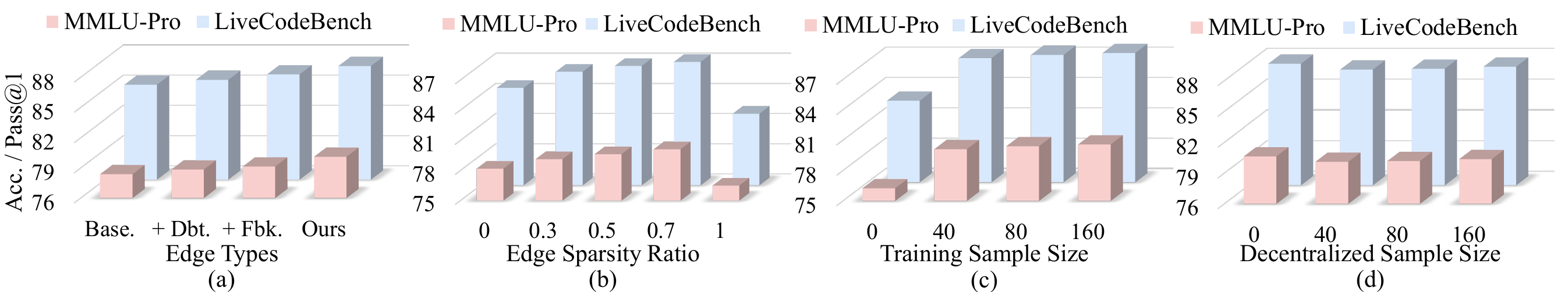} % 这里的图片文件名换成你自己的
  \caption{Dependence on hyperparameter and architecture designs. (a) Edge diversity vs. accuracy/pass@1, (b) edge sparsity vs. accuracy/pass@1, (c) training sample size vs. accuracy/pass@1, and (d) decentralized sample size vs. accuracy/pass@1.}
  \label{fig:analysis}
\end{figure*}

\noindent \textbf{Heterogeneous agents adaptability.} To validate the adaptability of \textsc{TopoDIM} within a heterogeneous MAS, we construct a collaborative framework comprising agents with varying capabilities: three \texttt{GPT-OSS-20B} agents and two \texttt{GPT-OSS-120B} agents, with the latter serves as the final decision-maker. As shown in Table~\ref{table2}, the results demonstrate that \textsc{TopoDIM} achieves SOTA performance across all datasets, yielding an average improvement of 1.86\% $\uparrow$ over existing homogeneous topology frameworks. \textsc{TopoDIM} leverages its adaptive sparsification design to effectively prune agents that make negligible contributions, thereby optimizing communication efficiency. A more comprehensive evaluation is provided in Appendix~\ref{sec:heter-ada}.

% \noindent \textbf{Heterogeneity adaptation.}
% % : \texttt{GPT-OSS-20B}, and \texttt{GPT-OSS-120B}. We
% To validate the adaptability of \textsc{TopoDIM} in handling tasks with a heterogeneous MAS, we comprise different language models of varying capabilities to establish a collaborative framework including three \texttt{GPT-OSS-20B} agents and two \texttt{GPT-OSS-120B} agents, designating \texttt{GPT-OSS-120B} as the final decision-maker. As detailed in Table~\ref{table2}, the data unequivocally confirms that \textsc{TopoDIM} establishes new SOTA performance across all types of datasets, registering an average improvement of 1.86\%$\uparrow$  over existing state-of-the-art homogeneous topology frameworks. A more comprehensive evaluation is detailed in Appendix~\ref{sec:heter-ada}.

\subsection{Framework Analysis}
We conduct comprehensive experiments using \texttt{GPT-OSS:120B} to validate the framework design of our proposed \textsc{TopoDIM}.

% \lm{explain little bit on how the experiments are performed. For example in Figure 5 it is not clear if you add new edge types incrementally, or one-by-one to baseline. }
\noindent \textbf{Edge diversity.}
% he influence of edge diversity on the collaborative problem-solving capacity of our MAS is presented in Figure~\ref{fig:analysis}(a), where \texttt{Base.} indicates only equipped with conditioned edges while \texttt{+ Fbk./Dbt.} signified adding feedback/debate edge types from \texttt{Base.} incrementally.
Figure~\ref{fig:analysis}(a) demonstrates the impact of edge diversity on the collaborative problem-solving capabilities of \textsc{TopoDIM}. Specifically, \texttt{Base.} denotes the configuration with only conditioned edges, while \texttt{+ Fbk./Dbt.} corresponds to the incremental addition of feedback and debate edge types.
Our systematic investigation reveals a positive correlation between edge diversity and performance. Notably, integrating all available edge types improves the prediction accuracy/pass@1 by 2.21\%$\uparrow$ on MMLU-Pro and 2.18\%$\uparrow$ on LiveCodeBench, respectively.
% This result substantiates the claim that a richer set of interaction modalities, represented by diverse edges, is crucial for fostering sophisticated agent collaboration on challenging tasks.

\noindent \textbf{Edge sparsity.} The amount of edges, controlled by the TopK sparsity budget ratio $\alpha \in \{0, 0.3, 0.5, 0.7, 1.0\}$, significantly impacts performance, as depicted in Figure~\ref{fig:analysis}(b).
The optimal accuracy/pass@1 (80.11\% on MMLU-Pro and 87.28\% on LiveCodeBench) is achieved at $\alpha = 0.7$, which reveals a crucial trade-off: a dense graph risks in creating redundant communication channels, leading to model hallucination and ultimately impairing decision-making performance.

\noindent \textbf{Training sample size.} We evaluate the data efficiency of \textsc{TopoDIM} by varying the training sample size $M$ (Figure~\ref{fig:analysis}(c)). The results demonstrate a monotonic improvement in performance as the number of samples increases, with the most significant gains occurring within the first 40 samples.%Crucially, the steepest performance increase occurr within the initial 40 samples, underscoring the remarkable data efficiency of \textsc{TopoDIM}.

\noindent \textbf{Decentralized sample size.} To evaluate the effectiveness of the distillation process, we vary the number of decentralized samples $M'$(from 0 to 160, where 0 indicates the direct use of the centralized policy to generate topology). As illustrated in Figure~\ref{fig:analysis}(d), \textsc{TopoDIM} achieves robust performance with as few as 40 samples, demonstrating the efficiency of the knowledge transfer.

\begin{table}[!t]
  \centering
  \caption{Performance comparison of heterogeneous agents distinguished by skill in MAS.}
  \vspace{-0.1em}
  \label{table2}
  \renewcommand\tabcolsep{5.3pt}
  \renewcommand\arraystretch{1.1}

  \resizebox{\linewidth}{!}{
    \begin{tabular}{l|ccccc}
      \Xhline{1.2pt}
      \rowcolor{CadetBlue!20}
      {\textbf{Method}} & \textbf{LiveCodeBench}& \textbf{MMLU-Pro} &\textbf{AIME} \\
      \Xhline{1.2pt}
      \multicolumn{4}{c}{\textit{Mode: Heterogeneous Skilled LLMs}} \\
      \hline

      Vanilla & 81.58 & 74.49 & 74.77\\

      \hdashline

      AgentDropout&  82.92\red{1.34} &  77.35\red{2.86} & 76.71\red{1.94}\\

      G-Designer&  83.86\red{2.28}& 78.18\red{3.69} & 77.10\red{2.32}\\

      \rowcolor{lightpurple}  \textbf{\textsc{TopoDIM}} & \textbf{84.52}\red{2.94} & \textbf{{79.52}\red{5.03}} &  \textbf{79.39}\red{4.62}\\

      \Xhline{1.2pt}
    \end{tabular}
  }
\end{table}

\begin{table}[!t]
  \centering
  \caption{Ablation study on the impact of topology and optimization strategy.}
  \vspace{-0.1em}
  \label{table:abaltion}
  \renewcommand\tabcolsep{5.3pt}
  \renewcommand\arraystretch{1.1}

  \resizebox{\columnwidth}{!}{
    \begin{tabular}{l|ccc}
      \Xhline{1.2pt}
      \rowcolor{CadetBlue!20}
      {\textbf{Method}} & \textbf{LiveCodeBench} & \textbf{MMLU-Pro} & \textbf{AIME} \\
      \Xhline{1.2pt}
      \hline

      \rowcolor{lightpurple}  \textbf{\textsc{TopoDIM}} & 87.28 & 80.11 & 80.34 \\

      \hdashline

      \textit{\textbf{w/ Rand}} & 82.52\blue{4.76}&76.36\blue{3.75}&75.71\blue{4.63} \\

      \textit{\textbf{w/o Graph}} & 85.49\blue{1.79}&78.62\blue{1.49}&78.77\blue{1.57} \\

      \textit{\textbf{w/o Baseline}}& 86.57\blue{0.71}&79.45\blue{0.66}&79.53\blue{0.81} \\

      \textit{\textbf{w/o ER}} & 86.42\blue{0.86}&78.97\blue{1.14}&78.95\blue{1.39} \\

      \Xhline{1.2pt}
    \end{tabular}
  }
\end{table}

% To validate the efficacy of our \textbf{\texttt{heterogeneous edge topology strategy}}, we test where heterogeneous edges were generated randomly \ding{182} \textbf{\textit{w/ Rand}} and replace $\mathcal{G}_{pri}$ with a fully connected graph \ding{183} \textbf{\textit{w/o Graph}}.
\noindent \textbf{Ablation studies.} To analyze \textsc{TopoDIM}'s architecture and validate the necessity of its core components, we perform a series of ablation studies.
First, to evaluate our \textbf{\texttt{heterogeneous interactions topology design}}, we consider two configurations: \ding{182} \textbf{\textit{w/ Rand}}, which features randomly generated heterogeneous edges, and \ding{183} \textbf{\textit{w/o Graph}}, which replaces $\mathcal{G}_\text{pri}$ with a fully connected graph (using random edge types). These modifications result in average performance drops of 5.60\%$\downarrow$ and 1.99\%$\downarrow$, respectively, underscoring the significance of the graph prior and our topology design.
Second, to evaluate the \textbf{\texttt{topology optimization strategy}}, we examine the contributions of the baseline and entropy regularization in optimization phases. Removing the baseline $b$: \ding{184} \textbf{\textit{w/o Baseline}}, which enhances training stability, leads to an accuracy decrease of 0.89\%$\downarrow$. Similarly, removing the entropy regularization term: \ding{185} \textbf{\textit{w/o ER}}, which facilitates policy exploration, causes accuracy to fall by 1.38\% $\downarrow$. These studies demonstrate the critical role of each component in \textsc{TopoDIM}.

\begin{figure}
  \centering
  \includegraphics[width=1\linewidth]{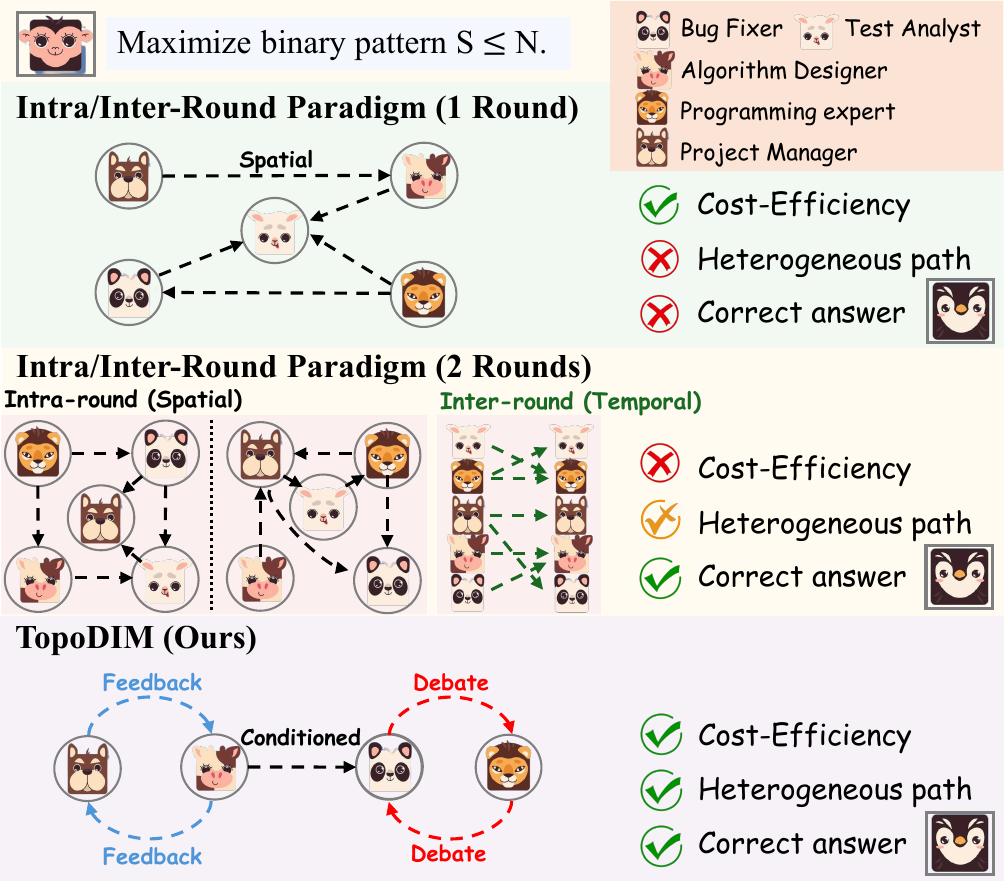}
  \caption{\textsc{TopoDIM} vs. Intra/inter-round dialogues methods on LiveCodeBench. Collaboration with diversity modes synthesizes various perspectives to ensure robust solutions with cost-efficiency.}
  \label{fig:case}
\end{figure}

\subsection{Case study}
We conduct a case study on LiveCodeBench using \texttt{GPT-OSS-120B} to evaluate the performance and optimized communication structures of \textsc{TopoDIM} against existing methods. Figure \ref{fig:case} visualizes the topologies generated by different approaches for a representative query.
The top panel depicts a typical intra-round dialogue graph. When handling complex tasks, such methods are often prone to information loss, leading to suboptimal performance in ambiguous scenarios.
The middle panel illustrates an intra/inter-round paradigm that facilitates agent collaboration through multi-turn dialogues. However, this approach incurs additional token overhead and increases susceptibility to hallucination.
In contrast, \textsc{TopoDIM} (bottom panel) dynamically constructs a sparse yet informative topology. By selectively activating conditional, feedback, and debate interaction patterns, it effectively preserves critical context without unnecessary token consumption, thereby successfully resolving complex reasoning challenges.
% Collaboration details are shown in Appendix~\ref{sec:case-study}.
Detailed dialogues are provided in Appendix~\ref{sec:case-study}.

\section{Related Work}
\subsection{LLM-based MAS}
Multi-agent systems(MAS) catalyze a paradigm shift from single-agent strategies \cite{yao2022react,zhang2026expseek} to collective intelligence for complex problems \cite{gong2025agents,ma2026talk2image,long2026emomas}. Pioneering frameworks demonstrated the efficacy of communicative role-playing and social simulation \cite{li2023camel,hu2026lying}, while task-oriented architectures have formalized collaborative workflows in graph generation \cite{zhang2026recont}, software engineering \cite{hong2023metagpt}, and logical reasoning \cite{chen2023agentverse} by integrating dynamic role assignment \cite{ye2025x}, iterative debate \cite{liang2023encouraging,zeng2025s,choi2025debate}, and shared memory \cite{wang2025mirix,zhang2025g3}. Additionally, recent work has introduced benchmarks for evaluating MAS \cite{zhang2026silo}. Recent advancements have corroborated decentralized architectures \cite{yang2025agentnetdecentralizedevolutionarycoordination} and diversified interaction patterns \cite{xue2025comas} to improve task proficiency and adaptability.
\textsc{TopoDIM} orchestrates decentralized structure and diverse interactions, demonstrating superior performance.

\subsection{Topology Design of MAS}
Communication topologies in LLM-based MAS are generally categorized into intra-round, inter-round, and hybrid structures \cite{zhang2024cut}. Early research primarily organizes either intra- or inter-round dialogues, including chain \cite{qian2024chatdev,holt2025l2maclargelanguagemodel}, tree \cite{wu2023autogen}, layered \cite{du2023improving,qian2024scaling}, and filtered graphs \cite{zheng2023progressive,zhuge2024gptswarm}. Recent advancements focus on generating task-specific adaptive topologies by integrating both intra- and inter-round interactions\cite{wang2025anymac,zhou2025multi,li2025assemble,jiang2025dynamic,wu2026st}. For instance, G-Designer \cite{zhang2025g} employs variational autoencoders to generate adaptive graph structures. While recent methods attempt to mitigate these costs via edge pruning \cite{zhang2024cut} and dropout \cite{wang2025agentdropout}, \textsc{TopoDIM} eliminates redundant dialogue rounds while maintaining diverse interactions, making a promising performance.
% achieves fundamental cost-efficiency by
\section{Conclusion}
In this paper, we aim at solving the structural redundancy of the hybrid intra/inter-round dialogue paradigm in MAS. We introduce \textsc{TopoDIM}, a novel decentralized framework for one-shot heterogeneous topology generation tailored adaptively to task queries.
% \textsc{TopoDIM} assign decisions of diverse interaction modes to each agent, establishing optimal multi-relationship network topology while reducing token consumption.
Experiments show that \textsc{TopoDIM} achieves SOTA performance in overall performance and cost-efficiency. Meanwhile, its adaptability to heterogeneous agent types allows to avoid the bucket effect. We hope our work will inspire future research on exploring synergy between cognitive mechanisms and interaction topology for scalable collective intelligence.

% Bibliography entries for the entire Anthology, followed by custom entries
%\bibliography{anthology,custom}
% Custom bibliography entries only
\section*{Limitations}
\label{sec:limitation}
\noindent \textbf{Interaction constraints.} While \textsc{TopoDIM} demonstrates competitive performance through conditioning, feedback, and debate mechanisms, our current focus prioritizes these direct interaction forms to maintain an optimal balance between predictive accuracy and token efficiency. Consequently, the scope of this work does not extend to complex organizational methods, such as dynamic coalition formation, where subsets of agents spontaneously align to address specific sub-problems.
% \noindent \textbf{\lm{***} Memory limitations.} \textsc{TopoDIM} relies on context engineering to leverage reasoning capabilities of LLMs within a simplified framework . We acknowledge that while \textsc{TopoDIM} mitigates noise and hallucinations in specific tasks, the absence of an integrated long-term memory module (e.g., vector databases) may limit the agents' ability to maintain cross-session consistency. \lm{We do not consider RAG, which might limit ****. }

\noindent \textbf{Heterogeneous agents adaptability.} We demonstrate \textsc{TopoDIM} outperforms baselines in heterogeneous agent settings, nevertheless finding the combination of high-performance and lightweight LLMs is non-trivial. In practice, we observed that simply introducing random agents yields small performance gains when tasks observably exceed agents capabilities. Furthermore, the deployment of heterogeneous LLMs imposes infrastructure challenges, as serving models with varying architectures and weights requires engineering support.

\noindent \textbf{Computational Overhead.} Compared to single-LLM approaches, MAS inevitably incurs more token consumption and latency due to extensive inter-agent communication~\cite{hu2026tapas}. While recent works attempt to mitigate this by organizing cooperation within the latent space~\cite{zou2025latent,fu2025cache}, we adhere to a topology-based paradigm to ensure compatibility with proprietary models. Future work will explore effective mechanisms to further optimize cost-efficiency while improving performance.

\noindent \textbf{}
% 1. more action in edge type
% 2. advanced memory mechansims management
% 3. bushu real-word constraints, we should allow the LLM heterogeouity if LLM difference how to cooperation and Governance
% 4. latent space cooperation less token fast future work agent 2 agent

\section*{Ethical Considerations}
\label{sec:ethical}
% we leverage existing LLMs to construct a multi-agent collaboration framework. W
% In this work, we rely on publicly available datasets and benchmarks, ensuring no private or personally identifiable information involved in our training or evaluation processes. Although we did not observe emergent harmful behaviors or unintended autonomous actions during our evaluation, we emphasize LLM-based multi-agent systems  may exhibit risks such as bias propagation or misleading information if not properly monitored. We explicitly state our framework serves as a coordination layer to enhance cooperative efficiency, without modifying the raw outputs or bypassing the existing safety guardrails of the base LLMs. For the AI utilization, we utilized AI assistants for grammatical checking and rephrasing.
This work relies exclusively on publicly available datasets and benchmarks, ensuring that no private or personally identifiable information is involved. While we did not observe emergent harmful behaviors or unintended autonomous actions during evaluation, we acknowledge that LLM-based multi-agent systems may inherit risks such as bias propagation or hallucinations. Crucially, our proposed framework serves as a coordination layer designed to enhance cooperative efficiency; it operates within the safety boundaries of the base LLMs without bypassing their existing guardrails. Finally, AI assistants were employed solely for grammatical error correction and rephrasing.

\section*{Acknowledgement}
This work was supported by the National Natural Science Foundation of China (T2293771), the STI 2030–Major Projects (2022ZD0211400), and National Natural Science Foundation of China (No. 42230406).

\bibliography{custom}

@article{qian2024scaling,
  title={Scaling large language model-based multi-agent collaboration},
  author={Qian, Chen and Xie, Zihao and Wang, Yifei and Liu, Wei and Zhu, Kunlun and Xia, Hanchen and Dang, Yufan and Du, Zhuoyun and Chen, Weize and Yang, Cheng and others},
  journal={arXiv preprint arXiv:2406.07155},
  year={2024}
}

@article{wang2025agentdropout,
  title={Agentdropout: Dynamic agent elimination for token-efficient and high-performance llm-based multi-agent collaboration},
  author={Wang, Zhexuan and Wang, Yutong and Liu, Xuebo and Ding, Liang and Zhang, Miao and Liu, Jie and Zhang, Min},
  journal={arXiv preprint arXiv:2503.18891},
  year={2025}
}

@article{zhang2025g,
  title={G-designer: Architecting multi-agent communication topologies via graph neural networks},
  author={Zhang, Guibin and Yue, Yanwei and Sun, Xiangguo and Wan, Guancheng and Yu, Miao and Fang, Junfeng and Wang, Kun and Chen, Tianlong and Cheng, Dawei},
  journal={arXiv preprint arXiv:2410.11782},
  year={2024}
}

@article{xue2025comas,
  title={CoMAS: Co-Evolving Multi-Agent Systems via Interaction Rewards},
  author={Xue, Xiangyuan and Zhou, Yifan and Zhang, Guibin and Zhang, Zaibin and Li, Yijiang and Zhang, Chen and Yin, Zhenfei and Torr, Philip and Ouyang, Wanli and Bai, Lei},
  journal={arXiv preprint arXiv:2510.08529},
  year={2025}
}

@inproceedings{du2023improving,
  title={Improving factuality and reasoning in language models through multiagent debate},
  author={Du, Yilun and Li, Shuang and Torralba, Antonio and Tenenbaum, Joshua B and Mordatch, Igor},
  booktitle={Forty-first International Conference on Machine Learning},
  year={2023}
}

@article{li2023camel,
  title={Camel: Communicative agents for" mind" exploration of large language model society},
  author={Li, Guohao and Hammoud, Hasan and Itani, Hani and Khizbullin, Dmitrii and Ghanem, Bernard},
  journal={Advances in Neural Information Processing Systems},
  volume={36},
  pages={51991--52008},
  year={2023}
}

@article{liang2023encouraging,
  title={Encouraging divergent thinking in large language models through multi-agent debate},
  author={Liang, Tian and He, Zhiwei and Jiao, Wenxiang and Wang, Xing and Wang, Yan and Wang, Rui and Yang, Yujiu and Shi, Shuming and Tu, Zhaopeng},
  journal={arXiv preprint arXiv:2305.19118},
  year={2023}
}

@article{choi2025debate,
  title={Debate or Vote: Which Yields Better Decisions in Multi-Agent Large Language Models?},
  author={Choi, Hyeong Kyu and Zhu, Xiaojin and Li, Yixuan},
  journal={arXiv preprint arXiv:2508.17536},
  year={2025}
}

@inproceedings{schlichtkrull2018modeling,
  title={Modeling relational data with graph convolutional networks},
  author={Schlichtkrull, Michael and Kipf, Thomas N and Bloem, Peter and Van Den Berg, Rianne and Titov, Ivan and Welling, Max},
  booktitle={European semantic web conference},
  pages={593--607},
  year={2018},
  organization={Springer}
}

@inproceedings{wang2024llm,
  title={Llm-enhanced cascaded multi-level learning on temporal heterogeneous graphs},
  author={Wang, Fengyi and Zhu, Guanghui and Yuan, Chunfeng and Huang, Yihua},
  booktitle={Proceedings of the 47th International ACM SIGIR Conference on Research and Development in Information Retrieval},
  pages={512--521},
  year={2024}
}

@article{zhang2024cut,
  title={Cut the crap: An economical communication pipeline for llm-based multi-agent systems},
  author={Zhang, Guibin and Yue, Yanwei and Li, Zhixun and Yun, Sukwon and Wan, Guancheng and Wang, Kun and Cheng, Dawei and Yu, Jeffrey Xu and Chen, Tianlong},
  journal={arXiv preprint arXiv:2410.02506},
  year={2024}
}

@article{williams1992simple,
  title={Simple statistical gradient-following algorithms for connectionist reinforcement learning},
  author={Williams, Ronald J},
  journal={Machine learning},
  volume={8},
  number={3},
  pages={229--256},
  year={1992},
  publisher={Springer}
}

@article{chen2024decentralized,
  title={Decentralized natural policy gradient with variance reduction for collaborative multi-agent reinforcement learning},
  author={Chen, Jinchi and Feng, Jie and Gao, Weiguo and Wei, Ke},
  journal={Journal of Machine Learning Research},
  volume={25},
  number={172},
  pages={1--49},
  year={2024}
}

@article{hendrycks2020measuring,
  title={Measuring massive multitask language understanding},
  author={Hendrycks, Dan and Burns, Collin and Basart, Steven and Zou, Andy and Mazeika, Mantas and Song, Dawn and Steinhardt, Jacob},
  journal={arXiv preprint arXiv:2009.03300},
  year={2020}
}

@article{cobbe2021training,
  title={Training verifiers to solve math word problems},
  author={Cobbe, Karl and Kosaraju, Vineet and Bavarian, Mohammad and Chen, Mark and Jun, Heewoo and Kaiser, Lukasz and Plappert, Matthias and Tworek, Jerry and Hilton, Jacob and Nakano, Reiichiro and others},
  journal={arXiv preprint arXiv:2110.14168},
  year={2021}
}

@article{roy2016solving,
  title={Solving general arithmetic word problems},
  author={Roy, Subhro and Roth, Dan},
  journal={arXiv preprint arXiv:1608.01413},
  year={2016}
}

@article{agarwal2025gpt,
  title={gpt-oss-120b \& gpt-oss-20b model card},
  author={Agarwal, Sandhini and Ahmad, Lama and Ai, Jason and Altman, Sam and Applebaum, Andy and Arbus, Edwin and Arora, Rahul K and Bai, Yu and Baker, Bowen and Bao, Haiming and others},
  journal={arXiv preprint arXiv:2508.10925},
  year={2025}
}

@article{wei2022chain,
  title={Chain-of-thought prompting elicits reasoning in large language models},
  author={Wei, Jason and Wang, Xuezhi and Schuurmans, Dale and Bosma, Maarten and Xia, Fei and Chi, Ed and Le, Quoc V and Zhou, Denny and others},
  journal={Advances in neural information processing systems},
  volume={35},
  pages={24824--24837},
  year={2022}
}

@inproceedings{chen2023agentverse,
  title={Agentverse: Facilitating multi-agent collaboration and exploring emergent behaviors},
  author={Chen, Weize and Su, Yusheng and Zuo, Jingwei and Yang, Cheng and Yuan, Chenfei and Chan, Chi-Min and Yu, Heyang and Lu, Yaxi and Hung, Yi-Hsin and Qian, Chen and others},
  booktitle={The Twelfth International Conference on Learning Representations},
  year={2023}
}

@inproceedings{zhuge2024gptswarm,
  title={Gptswarm: Language agents as optimizable graphs},
  author={Zhuge, Mingchen and Wang, Wenyi and Kirsch, Louis and Faccio, Francesco and Khizbullin, Dmitrii and Schmidhuber, J{\"u}rgen},
  booktitle={Forty-first International Conference on Machine Learning},
  year={2024}
}

@article{jiang2025dynamic,
  title={Dynamic Generation of Multi-LLM Agents Communication Topologies with Graph Diffusion Models},
  author={Jiang, Eric Hanchen and Wan, Guancheng and Yin, Sophia and Li, Mengting and Wu, Yuchen and Liang, Xiao and Li, Xinfeng and Sun, Yizhou and Wang, Wei and Chang, Kai-Wei and others},
  journal={arXiv preprint arXiv:2510.07799},
  year={2025}
}

@article{wang2020minilm,
  title={Minilm: Deep self-attention distillation for task-agnostic compression of pre-trained transformers},
  author={Wang, Wenhui and Wei, Furu and Dong, Li and Bao, Hangbo and Yang, Nan and Zhou, Ming},
  journal={Advances in neural information processing systems},
  volume={33},
  pages={5776--5788},
  year={2020}
}

@article{yang2025agentnetdecentralizedevolutionarycoordination,
      title={AgentNet: Decentralized Evolutionary Coordination for LLM-based Multi-Agent Systems}, 
      author={Yingxuan Yang and Huacan Chai and Shuai Shao and Yuanyi Song and Siyuan Qi and Renting Rui and Weinan Zhang},
      journal={arXiv preprint arXiv:2504.00587},
      year={2025},
}

@inproceedings{liu2024dynamic,
  title={A dynamic llm-powered agent network for task-oriented agent collaboration},
  author={Liu, Zijun and Zhang, Yanzhe and Li, Peng and Liu, Yang and Yang, Diyi},
  booktitle={First Conference on Language Modeling},
  year={2024}
}

@article{jain2024livecodebench,
  title={Livecodebench: Holistic and contamination free evaluation of large language models for code},
  author={Jain, Naman and Han, King and Gu, Alex and Li, Wen-Ding and Yan, Fanjia and Zhang, Tianjun and Wang, Sida and Solar-Lezama, Armando and Sen, Koushik and Stoica, Ion},
  journal={arXiv preprint arXiv:2403.07974},
  year={2024}
}

@inproceedings{hong2023metagpt,
  title={MetaGPT: Meta programming for a multi-agent collaborative framework},
  author={Hong, Sirui and Zhuge, Mingchen and Chen, Jonathan and Zheng, Xiawu and Cheng, Yuheng and Wang, Jinlin and Zhang, Ceyao and Wang, Zili and Yau, Steven Ka Shing and Lin, Zijuan and others},
  booktitle={The Twelfth International Conference on Learning Representations},
  year={2023}
}

@inproceedings{qian2024chatdev,
  title={Chatdev: Communicative agents for software development},
  author={Qian, Chen and Liu, Wei and Liu, Hongzhang and Chen, Nuo and Dang, Yufan and Li, Jiahao and Yang, Cheng and Chen, Weize and Su, Yusheng and Cong, Xin and others},
  booktitle={Proceedings of the 62nd Annual Meeting of the Association for Computational Linguistics (Volume 1: Long Papers)},
  pages={15174--15186},
  year={2024}
}

@article{zeng2025s,
  title={S$^2$-MAD: Breaking the Token Barrier to Enhance Multi-Agent Debate Efficiency},
  author={Zeng, Yuting and Huang, Weizhe and Jiang, Lei and Liu, Tongxuan and Jin, Xitai and Tiana, Chen Tianying and Li, Jing and Xu, Xiaohua},
  journal={arXiv preprint arXiv:2502.04790},
  year={2025}
}

@article{lei2024macm,
  title={Macm: Utilizing a multi-agent system for condition mining in solving complex mathematical problems},
  author={Lei, Bin and Zhang, Yi and Zuo, Shan and Payani, Ali and Ding, Caiwen},
  journal={Advances in Neural Information Processing Systems},
  volume={37},
  pages={53418--53437},
  year={2024}
}

@article{he2025llm,
  title={LLM-Based Multi-Agent Systems for Software Engineering: Literature Review, Vision, and the Road Ahead},
  author={He, Junda and Treude, Christoph and Lo, David},
  journal={ACM Transactions on Software Engineering and Methodology},
  volume={34},
  number={5},
  pages={1--30},
  year={2025},
  publisher={ACM New York, NY}
}

@article{islam2024mapcoder,
  title={Mapcoder: Multi-agent code generation for competitive problem solving},
  author={Islam, Md Ashraful and Ali, Mohammed Eunus and Parvez, Md Rizwan},
  journal={arXiv preprint arXiv:2405.11403},
  year={2024}
}

@article{ghareeb2025robin,
  title={Robin: A multi-agent system for automating scientific discovery},
  author={Ghareeb, Ali Essam and Chang, Benjamin and Mitchener, Ludovico and Yiu, Angela and Szostkiewicz, Caralyn J and Laurent, Jon M and Razzak, Muhammed T and White, Andrew D and Hinks, Michaela M and Rodriques, Samuel G},
  journal={arXiv preprint arXiv:2505.13400},
  year={2025}
}

@inproceedings{chen2025optima,
  title={Optima: Optimizing effectiveness and efficiency for llm-based multi-agent system},
  author={Chen, Weize and Yuan, Jiarui and Qian, Chen and Yang, Cheng and Liu, Zhiyuan and Sun, Maosong},
  booktitle={Findings of the Association for Computational Linguistics: ACL 2025},
  pages={11534--11557},
  year={2025}
}

@article{wu2023autogen,
  title={Autogen: Enabling next-gen llm applications via multi-agent conversation framework},
  author={Wu, Qingyun and Bansal, Gagan and Zhang, Jieyu and Wu, Yiran and Zhang, Shaokun and Zhu, Erkang and Li, Beibin and Jiang, Li and Zhang, Xiaoyun and Wang, Chi},
  journal={arXiv preprint arXiv:2308.08155},
  volume={3},
  number={4},
  year={2023}
}

@article{zheng2023progressive,
  title={Progressive-hint prompting improves reasoning in large language models},
  author={Zheng, Chuanyang and Liu, Zhengying and Xie, Enze and Li, Zhenguo and Li, Yu},
  journal={arXiv preprint arXiv:2304.09797},
  year={2023}
}

@article{scardamalia2006knowledge,
  title={Knowledge building},
  author={Scardamalia, Marlene and Bereiter, Carl},
  journal={The Cambridge},
  year={2006}
}

@article{wang2024mmlu,
  title={Mmlu-pro: A more robust and challenging multi-task language understanding benchmark},
  author={Wang, Yubo and Ma, Xueguang and Zhang, Ge and Ni, Yuansheng and Chandra, Abhranil and Guo, Shiguang and Ren, Weiming and Arulraj, Aaran and He, Xuan and Jiang, Ziyan and others},
  journal={Advances in Neural Information Processing Systems},
  volume={37},
  pages={95266--95290},
  year={2024}
}

@article{li2025assemble,
  title={Assemble your crew: Automatic multi-agent communication topology design via autoregressive graph generation},
  author={Li, Shiyuan and Liu, Yixin and Wen, Qingsong and Zhang, Chengqi and Pan, Shirui},
  journal={arXiv preprint arXiv:2507.18224},
  year={2025}
}

@article{liu2025deepseek,
  title={DeepSeek-V3. 2: Pushing the Frontier of Open Large Language Models},
  author={Liu, Aixin and Mei, Aoxue and Lin, Bangcai and Xue, Bing and Wang, Bingxuan and Xu, Bingzheng and Wu, Bochao and Zhang, Bowei and Lin, Chaofan and Dong, Chen and others},
  journal={arXiv preprint arXiv:2512.02556},
  year={2025}
}

@article{team2025gemma,
  title={Gemma 3 technical report},
  author={Team, Gemma and Kamath, Aishwarya and Ferret, Johan and Pathak, Shreya and Vieillard, Nino and Merhej, Ramona and Perrin, Sarah and Matejovicova, Tatiana and Ram{\'e}, Alexandre and Rivi{\`e}re, Morgane and others},
  journal={arXiv preprint arXiv:2503.19786},
  year={2025}
}

@inproceedings{yao2022react,
  title={React: Synergizing reasoning and acting in language models},
  author={Yao, Shunyu and Zhao, Jeffrey and Yu, Dian and Du, Nan and Shafran, Izhak and Narasimhan, Karthik R and Cao, Yuan},
  booktitle={The eleventh international conference on learning representations},
  year={2022}
}

@article{ye2025x,
  title={X-MAS: Towards Building Multi-Agent Systems with Heterogeneous LLMs},
  author={Ye, Rui and Liu, Xiangrui and Wu, Qimin and Pang, Xianghe and Yin, Zhenfei and Bai, Lei and Chen, Siheng},
  journal={arXiv preprint arXiv:2505.16997},
  year={2025}
}

@article{zhang2025g3,
  title={G-Memory: Tracing Hierarchical Memory for Multi-Agent Systems},
  author={Zhang, Guibin and Fu, Muxin and Wan, Guancheng and Yu, Miao and Wang, Kun and Yan, Shuicheng},
  journal={arXiv preprint arXiv:2506.07398},
  year={2025}
}

@article{wang2025mirix,
  title={Mirix: Multi-agent memory system for llm-based agents},
  author={Wang, Yu and Chen, Xi},
  journal={arXiv preprint arXiv:2507.07957},
  year={2025}
}

@article{holt2025l2maclargelanguagemodel,
      title={L2MAC: Large Language Model Automatic Computer for Extensive Code Generation}, 
      author={Samuel Holt and Max Ruiz Luyten and Mihaela van der Schaar},
      journal={arXiv preprint arXiv:2310.02003},
      year={2025},
}

@article{sutton1999policy,
  title={Policy gradient methods for reinforcement learning with function approximation},
  author={Sutton, Richard S and McAllester, David and Singh, Satinder and Mansour, Yishay},
  journal={Advances in neural information processing systems},
  volume={12},
  year={1999}
}

@misc{ollama,
  author = {Ollama},
  title = {Ollama: Large Language Model Runner},
  year = {2023},
  url = {https://ollama.com/},
  version = {0.12.10},
  note = {Software available at https://github.com/ollama/ollama}
}

@misc{openai2025gpt5,
  title={GPT-5 System Card},
  author={OpenAI},
  year={2025},
  howpublished={\url{https://openai.com/index/gpt-5-system-card/}},
  note={Accessed: 2025-08-13}
}

@article{zhou2025multi,
  title={Multi-agent design: Optimizing agents with better prompts and topologies},
  author={Zhou, Han and Wan, Xingchen and Sun, Ruoxi and Palangi, Hamid and Iqbal, Shariq and Vuli{\'c}, Ivan and Korhonen, Anna and Ar{\i}k, Sercan {\"O}},
  journal={arXiv preprint arXiv:2502.02533},
  year={2025}
}

@article{wang2025anymac,
  title={AnyMAC: Cascading Flexible Multi-Agent Collaboration via Next-Agent Prediction},
  author={Wang, Song and Tan, Zhen and Chen, Zihan and Zhou, Shuang and Chen, Tianlong and Li, Jundong},
  journal={arXiv preprint arXiv:2506.17784},
  year={2025}
}

@article{zou2025latent,
  title={Latent Collaboration in Multi-Agent Systems}, 
  author={Jiaru Zou and Xiyuan Yang and Ruizhong Qiu and Gaotang Li and Katherine Tieu and Pan Lu and Ke Shen and Hanghang Tong and Yejin Choi and Jingrui He and James Zou and Mengdi Wang and Ling Yang},
  journal={arXiv preprint arXiv:2511.20639},
  year={2025},
}

@misc{maa_aime,
  title = {American Invitational Mathematics Examination (AIME)},
  author = {{Mathematical Association of America}},
  year = {1983--2025},
  note = {Accessed via Art of Problem Solving or [Insert Dataset Name]},
  url = {https://www.maa.org/math-competitions/aime}
}

@article{zhang2026silo,
  title={Silo-Bench: A Scalable Environment for Evaluating Distributed Coordination in Multi-Agent LLM Systems},
  author={Zhang, Yuzhe and Liu, Feiran and Shan, Yi and Huang, Xinyi and Yang, Xin and Zhu, Yueqi and Cheng, Xuxin and Liu, Cao and Zeng, Ke and Zhang, Terry Jingchen and others},
  journal={arXiv preprint arXiv:2603.01045},
  year={2026}
}

@article{hu2026context,
  title={Context-Agent: Dynamic Discourse Trees for Non-Linear Dialogue},
  author={Hu, Junan and Guo, Shudan and Liu, Wenqi and Yin, Jianhua and Wei, Yinwei},
  journal={arXiv preprint arXiv:2604.05552},
  year={2026}
}

@article{zhang2026expseek,
  title={ExpSeek: Self-Triggered Experience Seeking for Web Agents},
  author={Zhang, Wenyuan and Zhang, Xinghua and Yu, Haiyang and Nie, Shuaiyi and Wu, Bingli and Yue, Juwei and Liu, Tingwen and Li, Yongbin},
  journal={arXiv preprint arXiv:2601.08605},
  year={2026}
}

@article{gong2025agents,
  title={Agents with foundation models: advance and vision},
  author={Gong, Chenghua and Li, Xiang},
  journal={Frontiers of Computer Science},
  volume={19},
  number={4},
  pages={194330},
  year={2025}
}

@article{ma2026tspo,
  title={TSPO: Breaking the Double Homogenization Dilemma in Multi-turn Search Policy Optimization},
  author={Ma, Shichao and Ma, Zhiyuan and Yang, Ming and Li, Xiaofan and Wu, Xing and Du, Jintao and Cheng, Yu and Wang, Weiqiang and Liu, Qiliang and Zhou, Zhengyang and others},
  journal={arXiv preprint arXiv:2601.22776},
  year={2026}
}

@inproceedings{ma2026talk2image,
  title={Talk2image: A multi-agent system for multi-turn image generation and editing},
  author={Ma, Shichao and Guo, Yunhe and Su, Jiahao and Huang, Qihe and Zhou, Zhengyang and Wang, Yang},
  booktitle={Proceedings of the AAAI Conference on Artificial Intelligence},
  volume={40},
  number={38},
  pages={32437--32445},
  year={2026}
}

@article{hu2026lying,
  title={Lying with Truths: Open-Channel Multi-Agent Collusion for Belief Manipulation via Generative Montage},
  author={Hu, Jinwei and Huang, Xinmiao and Sun, Youcheng and Dong, Yi and Huang, Xiaowei},
  journal={arXiv preprint arXiv:2601.01685},
  year={2026}
}

@inproceedings{hu2026tapas,
  title={Tapas are free! training-free adaptation of programmatic agents via llm-guided program synthesis in dynamic environments},
  author={Hu, Jinwei and Dong, Yi and Sun, Youcheng and Huang, Xiaowei},
  booktitle={Proceedings of the AAAI Conference on Artificial Intelligence},
  volume={40},
  number={35},
  pages={29477--29485},
  year={2026}
}

@article{liu2026bapo,
  title={BAPO: Boundary-Aware Policy Optimization for Reliable Agentic Search},
  author={Liu, Shiyu and Yin, Yongjing and Yan, Jianhao and Tang, Yunbo and Zhang, Qinggang and Li, Bei and Chen, Xin and Wang, Jingang and Cai, Xunliang and Su, Jinsong},
  journal={arXiv preprint arXiv:2601.11037},
  year={2026}
}

@article{fu2025cache,
  title={Cache-to-cache: Direct semantic communication between large language models},
  author={Fu, Tianyu and Min, Zihan and Zhang, Hanling and Yan, Jichao and Dai, Guohao and Ouyang, Wanli and Wang, Yu},
  journal={arXiv preprint arXiv:2510.03215},
  year={2025}
}

@article{wu2026st,
  title={ST-EVO: Towards Generative Spatio-Temporal Evolution of Multi-Agent Communication Topologies},
  author={Wu, Xingjian and Liu, Xvyuan and Lu, Junkai and Wang, Siyuan and Qiu, Xiangfei and Shu, Yang and Hu, Jilin and Guo, Chenjuan and Yang, Bin},
  journal={arXiv preprint arXiv:2602.14681},
  year={2026}
}

@article{zhang2026recont,
      title={ReContraster: Making Your Posters Stand Out with Regional Contrast}, 
      author={Peixuan Zhang and Zijian Jia and Ziqi Cai and Shuchen Weng and Si Li and Boxin Shi},
      journal={arXiv preprint arXiv:2604.10442},
      year={2026}
}

@article{long2026emomas,
  title={EmoMAS: Emotion-Aware Multi-Agent System for High-Stakes Edge-Deployable Negotiation with Bayesian Orchestration},
  author={Long, Yunbo and Liu, Yuhan and Xu, Liming},
  journal={arXiv preprint arXiv:2604.07003},
  year={2026}
}

% \newpage
\appendix

\section{Algorithm workflow}
The algorithm workflow of \textsc{TopoDIM} is illustrated in Algorithm~\ref{Algo:topodim}. Specifically, \textsc{TopoDIM} comprises two training stages. In stage 1, \textsc{TopoDIM}  iterates through all training queries and employs policy optimization to update the encoder and decoder parameters, incorporating structural constraints and adaptive sparsification. In stage 2, \textsc{TopoDIM} performs decentralized policy distillation using the same training set as in Stage 1, updating the parameters of lightweight networks which are deployed on each agent itself for a purpose of flexible and private decision for .
During the inference phase, \textsc{TopoDIM} generates a communication topology and traverses it in a breadth-first manner. By facilitating agent cooperation through three semantically rich interaction modes, the method achieves competitive performance on complex tasks.

\begin{figure}[H] % 注意这里是 figure*，[t] 表示置顶
  \centering
  % --- 第1张图 ---
  \includegraphics[width=\linewidth]{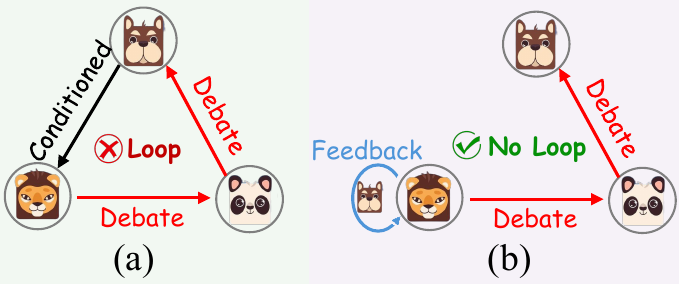}
  \caption{Illustration of acyclic property for three edge types, including (a) loop collaboration with conditioned and debate edges and (b) no loop collaboration with three edge types. The solid arrows only indicate communication directions.}
  \label{fig:edge}
\end{figure}

\label{sec:Algo-supp}

% This Appendix explains why the graph being acyclic relies on $\mathcal{R}_{\text{constr}} = {\mathrm{\{conditioned,debate\}}}$ but not $\mathcal{R}_{\text{free}} = {\mathrm{\{feedback\}}}$. According to Figure~\ref{fig:interaction} in the main text, the solution result from an interaction in $\mathcal{R}_{\text{constr}}$ will be retained for both source and target nodes, while for $\mathcal{R}_{\text{free}}$ only the source node will keep the final result. As a consequence,

\section{Acyclic Property}
\label{sec:edge-supplement}
We impose acyclic structural constraints exclusively on conditioned and debate edges, leaving feedback edges unconstrained. Specifically, structural constraints are applied exclusively to conditioned and debate edges, which must remain acyclic to preclude circular dependencies (Figure~\ref{fig:edge}(a)). In contrast, feedback edges are allowed for arbitrary connectivity (Figure~\ref{fig:edge}(b)). During traversal, feedback edges are prioritized for immediate execution: they are excluded from degree calculations and removed upon visit. Meanwhile other edge types adhere to standard breadth-first scheduling.
% The rationale lies in the directed nature of these interaction modes, where the initiator and finisher are distinct agents. The relation $r_{constr}$ is subject to acyclic structural constraints in Section~\ref{sec:M2}, precluding circular dependencies shown in Figure \ref{fig:edge}(a), where a closed communication loop would result in system deadlock. By contrast, $r_{free} = \mathrm{feedback}$ essentially forming a reflexive edge (starting and ending with the same agent), as depicted in Figure~\ref{fig:edge}(b). Since this edge type is defined to contribute zero in-degree during the breadth-first traversal, it does not introduce circular dependencies.

% \section{Execution Sequence Supplement}
% \label{sec:Exe-supplement}

\section{Data Statistics}
\label{sec:data-statis}
We summarized the data statistics in Table~\ref{table:data_stat}. In this paper we conduct experiments across three representative categories of datasets, incorporating challenging benchmarks such as MMLU-Pro, AIME, and LiveCodeBench to make a comprehensive evaluation of our proposed \textsc{TopoDIM}. For the AIME benchmark, given the limited volume of available problems, we utilize the entire collection of samples from 2023 to 2025 for training (40 samples) and evaluation (50 samples)~\cite{maa_aime}. For LiveCodeBench, we employ Release V1 version of for a moderately challenging evaluation on code generation tasks, which comprises data collected between May 2023 and March 2024. For MMLU-Pro, we adopt the same data processing pipeline as applied to MMLU in previous work~\cite{zhang2025g}, selecting 153 samples from the test set as our evaluation subset. The experimental settings for the remaining datasets, including MultiArith, HumanEval, and GSM8K, remain consistent with prior research~\cite{wang2025agentdropout} to ensure fair comparison.

\begin{table}[h]
  \centering
  \caption{Dataset descriptions and statistics.}
  \vspace{-0.1em}
  \label{table:data_stat}
  \renewcommand\tabcolsep{5.3pt}
  \renewcommand\arraystretch{1.1}

  \resizebox{\linewidth}{!}{
    \begin{tabular}{llccc}
      \Xhline{1.2pt}
      \rowcolor{CadetBlue!20}

      \rowcolor{CadetBlue!20}  \textbf{Category} & \textbf{Dataset} & \textbf{Answer Type} & \textbf{Metric} & \textbf{\#Test}  \\
      \Xhline{1.2pt}
      \multirow{1}{*}{General reasoning} & MMLU-Pro & Multi-choice & Acc. & 153  \\
      \midrule
      \multirow{3}{*}{Math reasoning}
      & AIME$_{23-25}$ & Number & Acc. & 50  \\
      & MultiArith & Number & Acc. & 600  \\
      & GSM8K & Number & Acc. & 1,319 \\
      \midrule
      \multirow{2}{*}{Code generation} & LiveCodeBench$_{\text{v}1}$ & Code & Pass@1 & 360  \\
      & HumanEval & Code & Pass@1 & 164  \\

      \Xhline{1.2pt}
    \end{tabular}
  }
\end{table}
% \section{Limitation}
% \label{sec:appendix D}

\section{Cost Efficiency}
\label{sec:cost-ef}
We provide a more detailed display for cost-efficiency of \textsc{TopoDIM}, shown in Table~\ref{table:consump}. \textsc{TopoDIM} cuts down the token expenditure, saving maximum prompt and completion token by 57.82\% and 22.05\% compared to the most efficient framework.
Moreover, we conducted experiments to evaluate GPU resource overhead during the inference phase. By randomly instantiating 1,000 nodes, we demonstrated that our decentralized design, where each agent operates a lightweight network, requires merely 6.36 GB of memory, achieving significant cost efficiency.

\begin{table}[]
  \centering
  \caption{Performance comparison of heterogeneous agents distinguished by sources and capacities in MAS.}
  \vspace{-0.1em}
  \label{table:heterosu}
  \renewcommand\tabcolsep{5.3pt}
  \renewcommand\arraystretch{1.1}

  \resizebox{\linewidth}{!}{
    \begin{tabular}{l|ccccc}
      \Xhline{1.2pt}
      \rowcolor{CadetBlue!20}
      {\textbf{Method}} & \textbf{LiveCodeBench}& \textbf{MMLU-Pro} &\textbf{AIME} \\
      \Xhline{1.2pt}
      \multicolumn{4}{c}{\textbf{Decision-maker:} \texttt{GPT-OSS:20B}}  \\
      \multicolumn{4}{c}{\textbf{Collabration:} 3 \texttt{Gemma3-it:12B}, 2 \texttt{GPT-OSS:20B}} \\

      \hline

      Vanilla & 70.40 & 68.84 & 29.57 \\

      \hdashline

      AgentDropout &  71.26\red{0.86} & 69.57\red{0.73} & 29.80\red{0.23} \\

      G-Designer &  71.18\red{0.78} & 70.46\red{1.62} & 29.24\blue{0.33} \\

      \rowcolor{lightpurple}  \textbf{\textsc{TopoDIM}} & \textbf{72.23}\red{1.83} & \textbf{{71.20}\red{2.36}} &  \textbf{30.18}\red{0.61}\\
      \hline
      \multicolumn{4}{c}{\textbf{Decision-maker:} \texttt{DeepSeek-V3.2}} \\
      \multicolumn{4}{c}{\textbf{Collabration:} 3 \texttt{GPT-OSS:20B}, 2 \texttt{DeepSeek-V3.2}} \\

      \hline
      Vanilla & 78.38 & 78.68 & 64.37\\

      \hdashline

      AgentDropout&  79.52\red{1.14} &  80.10\red{1.42} & 64.55\red{0.18}\\

      G-Designer &  80.17\red{1.79}& 80.68\red{2.00} & 65.23\red{0.86}\\

      \rowcolor{lightpurple}  \textbf{\textsc{TopoDIM}} & \textbf{82.46}\red{4.08} & \textbf{{82.37}\red{3.69}} &  \textbf{66.59}\red{2.22}\\
      \Xhline{1.2pt}

    \end{tabular}
  }
\end{table}

\begin{table*}[t]
  \centering
  \caption{Comparison of token consumption. Ctok. and Ptok. indicate prompt and complete token consumption.}
  \vspace{-0.1em}
  \label{table:consump}
  \renewcommand\tabcolsep{5.3pt}
  \renewcommand\arraystretch{1.1}

  \resizebox{\textwidth}{!}{
    \begin{tabular}{l|cccccccccc}
      \Xhline{1.2pt}
      \multirow{2}{*}{\textbf{Consumption}}&\multicolumn{2}{c}{\textbf{LLM-Debate}} & \multicolumn{2}{c}{\textbf{GPTSwarm}} & \multicolumn{2}{c}{\textbf{G-Designer}} & \multicolumn{2}{c}{\textbf{AgentDropout}}&\multicolumn{2}{c}{\textbf{\textsc{TopoDIM}}}  \\
      \cline{2-11}

      &\textbf{MMLU-Pro} & \textbf{LiveCodeBench} & \textbf{MMLU-Pro} & \textbf{LiveCodeBench}&\textbf{MMLU-Pro} & \textbf{LiveCodeBench} & \textbf{MMLU-Pro} & \textbf{LiveCodeBench} &\textbf{MMLU-Pro} & \textbf{LiveCodeBench}  \\
      \Xhline{1.2pt}
      \textbf{Ptok.} &7.31 M & 13.72 M & 6.14 M & 9.46 M & 3.93 M & 6.37 M & 2.09 M & 5.30 M & 0.88 M & 3.02 M  \\
      \textbf{Ctok.} &2.87 M & 4.63 M & 2.40 M & 3.15 M & 1.56 M & 2.21 M & 0.98 M & 1.85 M & 0.76 M & 1.69 M  \\
      \textbf{Total} &10.18 M & 18.36 M & 8.54 M & 12.61 M & 5.49 M & 8.58 M & 3.06 M & 7.16 M & 1.64 M & 4.71 M  \\

      \Xhline{1.2pt}
    \end{tabular}
  }
\end{table*}

\section{Heterogeneous Agents Adaptability}
\label{sec:heter-ada}
To provide a more comprehensive evaluation of \textsc{TopoDIM}'s adaptability to heterogeneous agents, we conducted experiments on two additional configurations involving agents with distinct sources and capacities. As detailed in Table~\ref{table:heterosu}, we combined \texttt{GPT-OSS:20B} with \texttt{DeepSeek-V3.2} and \texttt{Gemma3-it:12b}, respectively. In these settings, \textsc{TopoDIM} consistently outperformed other baselines across three representative datasets, achieving average improvements of 1.32\% and 1.15\%, respectively. It is worth noting that identifying effective combinations of high-performance and lightweight LLMs is non-trivial; naively introducing random agents with heterogeneous skills can sometimes be counterproductive (e.g., the performance degradation observed with G-Designer). In contrast, \textsc{TopoDIM} leverages its adaptive sparsification design to effectively prune agents that make negligible contributions, thereby optimizing communication efficiency.

% \section{Robustness Analysis}
% \label{sec:robu-ana}

\section{Robustness Analysis}
\begin{figure}[] % 注意这里是 figure*，[t] 表示置顶
  \centering
  % --- 第1张图 ---
  \includegraphics[width=\linewidth]{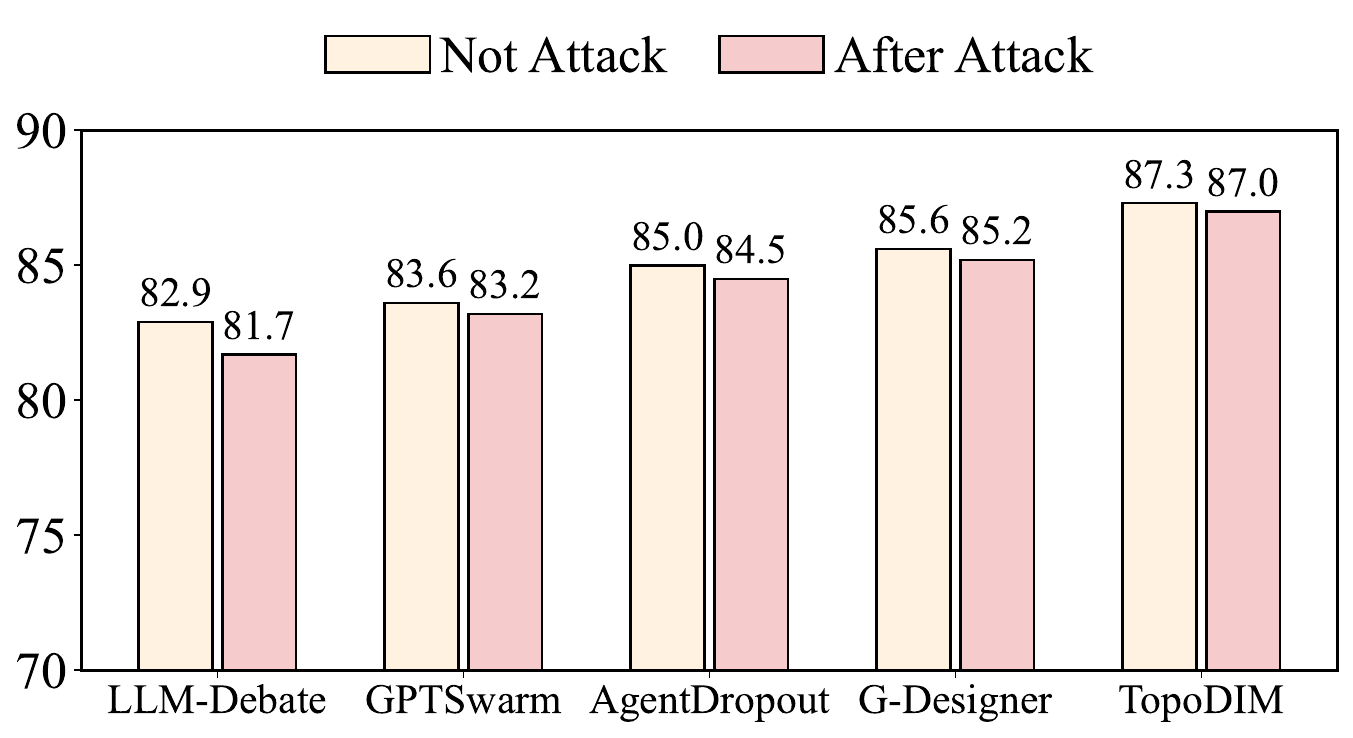}
  \caption{Pass@1 performance comparison of \textsc{TopoDIM} and other multi-agent baselines on LiveCodeBench pre- and post-prompt attacks. Dynamic topology frameworks exhibit competitive structural robustness, demonstrating small performance degradation.}
  \label{fig:robost}
\end{figure}

We evaluated the robustness of \textsc{TopoDIM} by conducting experiments on LiveCodeBench with \texttt{GPT-OSS:120B}. An agent is randomly selected and injected with prompts that drive it to generate failure outputs. As illustrated in Figure~\ref{fig:robost}, thanks to its heterogeneous adaptability, \textsc{TopoDIM} exhibits remarkable robustness to these attacks, limiting the performance degradation to 0.3 points. Conversely, LLM-Debate incurs a drop of 1.2 due to its fixed communication topology. Meanwhile, dynamic topology frameworks like GPTSwarm, G-Designer, and AgentDropout also display competitive structural robustness, with an average drop of 0.43. This adaptive capability is vital for identifying malicious agents, underscoring the superiority of dynamic topologies in maintaining structural integrity.

\section{Case Study}
\label{sec:case-study}
\begin{figure}
  \centering
  \includegraphics[width=1\linewidth]{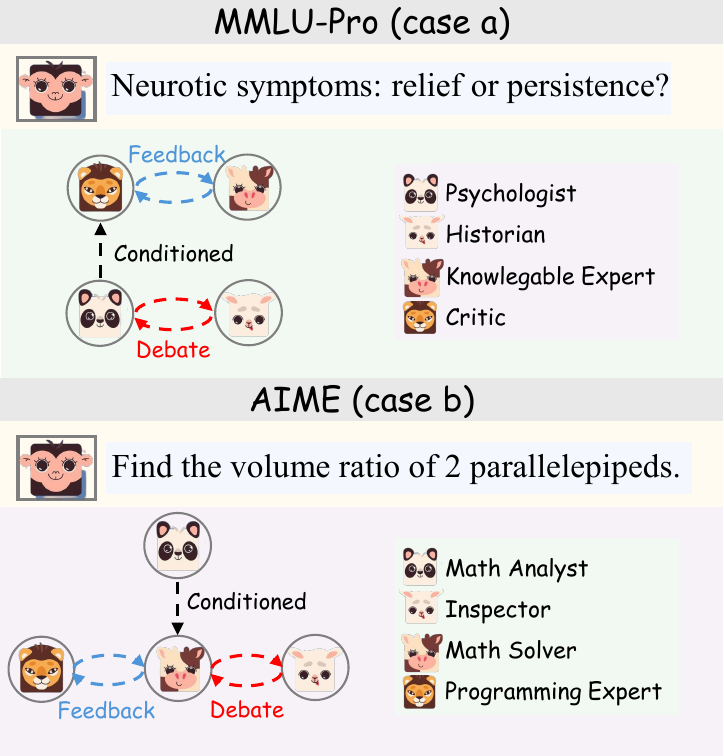}
  % \caption[Interaction modes of \textsc{TopoDIM}]{Interaction modes of \textsc{TopoDIM}. Information flow for $\{\mathrm{conditioned, debate}\}$ proceeds from \raisebox{-0.3em}{\includegraphics[height=1.3em]{figure/icon1.pdf}} to \raisebox{-0.3em}{\includegraphics[height=1.3em]{figure/icon2.pdf}}, while $\mathrm{feedback}$ starts and ends at \raisebox{-0.3em}{\includegraphics[height=1.3em]{figure/icon1.pdf}}.}
  \caption{Case studies of the communication topologies designed by \textsc{TopoDIM} on MMLU-Pro (case a) and AIME (case b) benchmarks.}
  \label{fig:cases-specific}
\end{figure}

\noindent \textbf{Task-specific Patterns.} We conduct case studies with \texttt{GPT-OSS-120B} on AIME (math) and MMLU-Pro (general) to illustrate task-specific collaboration patterns, as shown in Figure~\ref{fig:cases-specific}. For an MMLU-Pro psychology question, \textsc{TopoDIM} generates an efficient topology that coordinates multiple agents to solve the problem. The psychologist and historian first engage in a debate. The critic then reviews the psychologist's response and consults a knowledgeable expert for professional feedback. For an AIME problem, the math analyst first examines the question and passes the result to the math solver. Conditioned on this output, the math solver works out the solution and requests feedback from a programming expert for verification. Finally, the math solver and inspector enter a debate to further refine the solution.

\begin{figure*}[] % 注意这里是 figure*，[t] 表示置顶
  \centering
  % --- 第1张图 ---
  \includegraphics[width=0.9\linewidth]{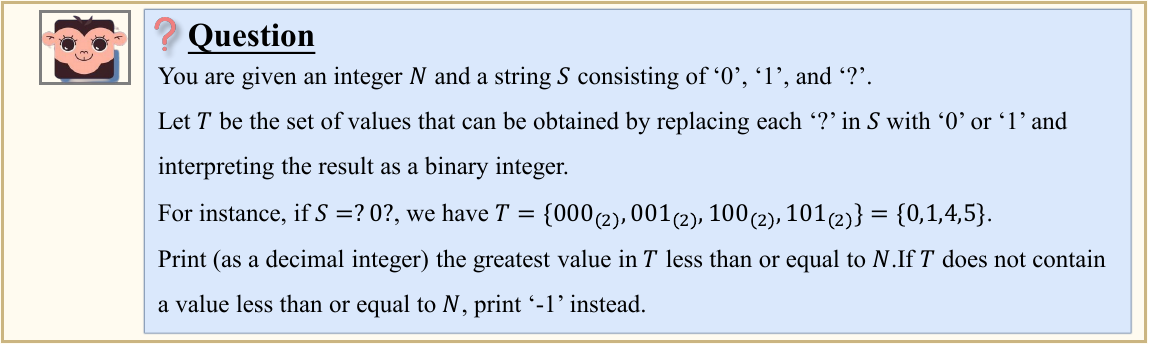}
  \caption{Question overview of case study.}
  \label{fig:case_d1}
\end{figure*}

\begin{figure*}[] % 注意这里是 figure*，[t] 表示置顶
  \centering
  % --- 第1张图 ---
  \includegraphics[width=0.9\linewidth]{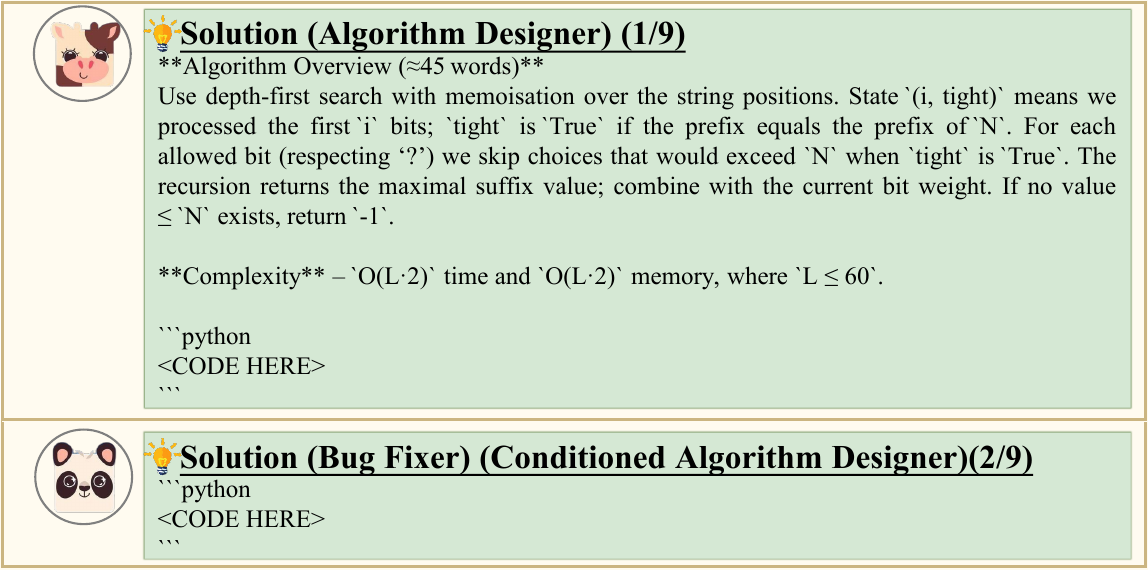}
  \caption{Detailed dialogues generated in the $\mathrm{Conditioned}$ mode.}
  \label{fig:case_d2}
\end{figure*}

\begin{figure*}[] % 注意这里是 figure*，[t] 表示置顶
  \centering
  % --- 第1张图 ---
  \includegraphics[width=0.9\linewidth]{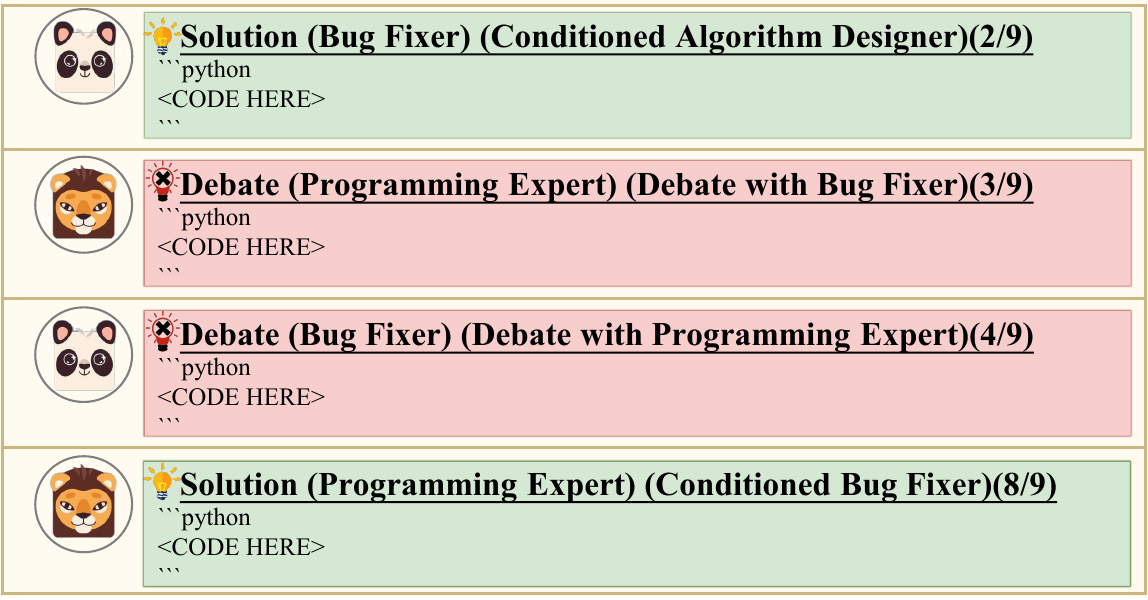}
  \caption{Detailed dialogues generated in the $\mathrm{Debate}$ mode.}
  \label{fig:case_d3}
\end{figure*}

\begin{figure*}[] % 注意这里是 figure*，[t] 表示置顶
  \centering
  % --- 第1张图 ---
  \includegraphics[width=0.9\linewidth]{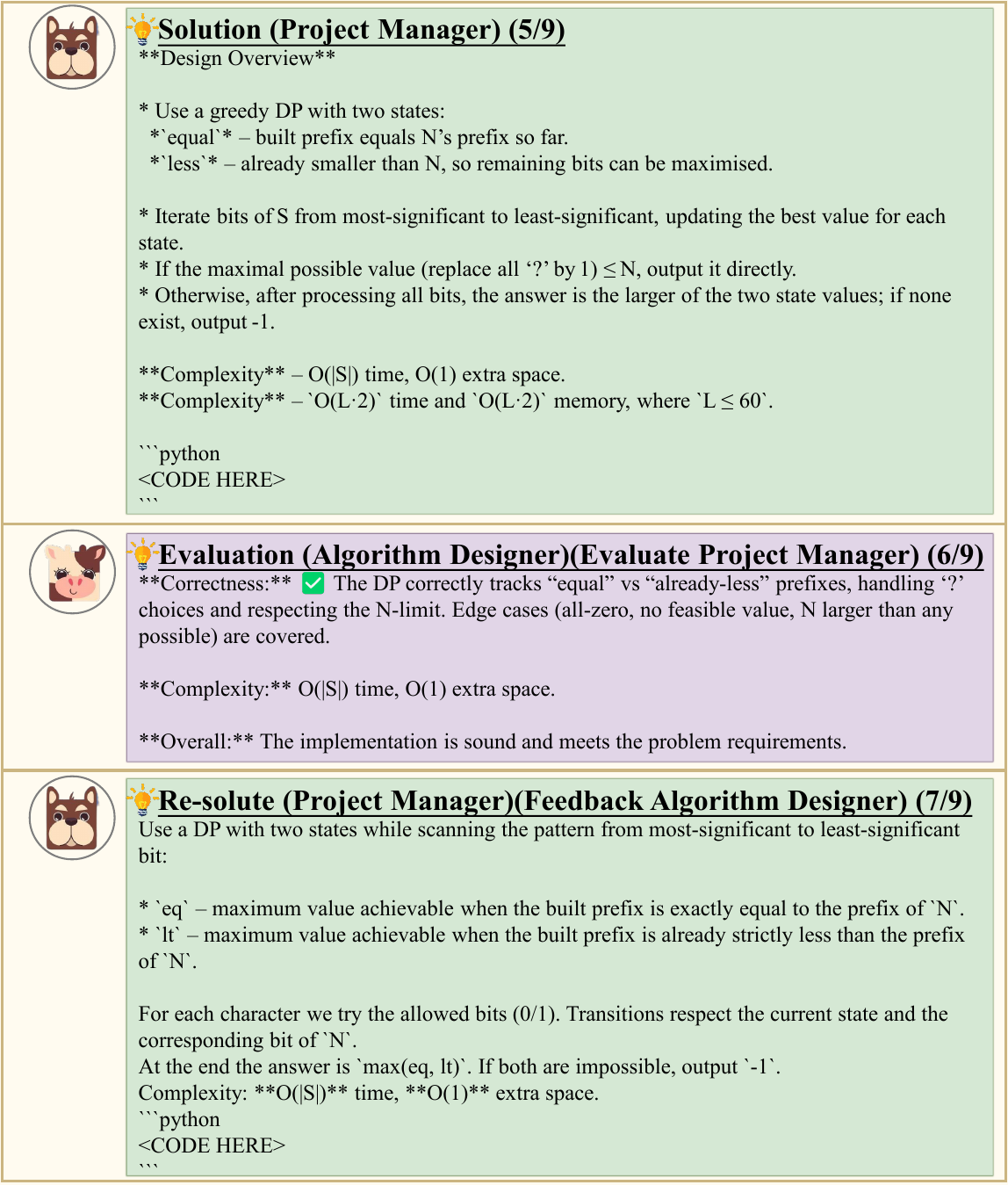}
  \caption{Detailed dialogues generated in the $\mathrm{Feedback}$ mode.}
  \label{fig:case_d4}
\end{figure*}

\begin{figure*}[] % 注意这里是 figure*，[t] 表示置顶
  \centering
  % --- 第1张图 ---
  \includegraphics[width=0.9\linewidth]{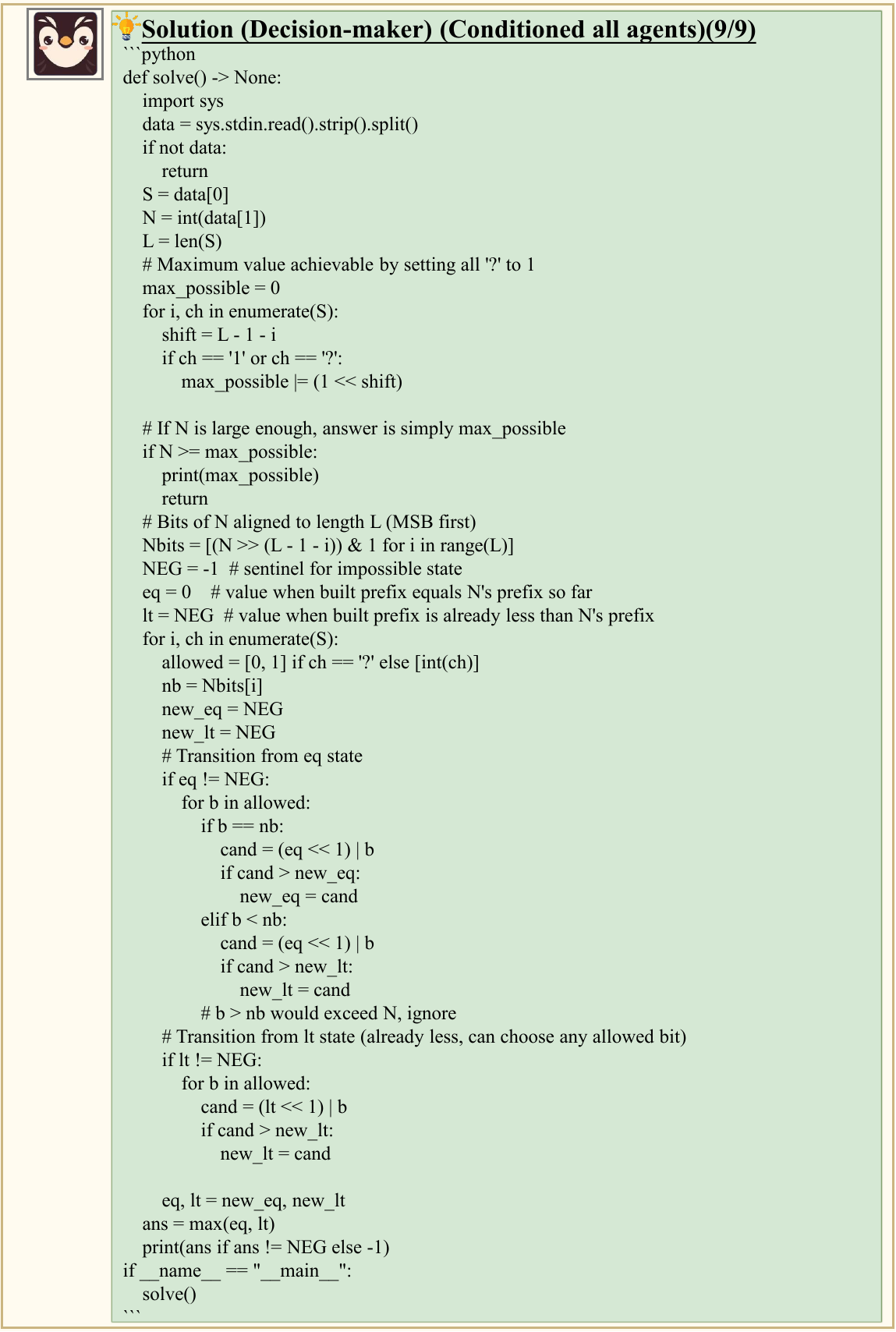}
  \caption{Detailed decision-making overview of case study.}
  \label{fig:case_d5}
\end{figure*}

\noindent \textbf{Specific Cases.}
While the main text briefly compares the communication sequence of \textsc{TopoDIM} with two typical intra- and inter-round dialogue paradigms using \texttt{GPT-OSS:120B} on a LiveCodeBench question, this section presents the detailed cooperative dialogues among different agents, as illustrated in Figures~\ref{fig:case_d1} through~\ref{fig:case_d5}.
Upon receiving a question $q$, \textsc{TopoDIM} effectively generates a communication graph to orchestrate agent collaboration. As shown in Figure~\ref{fig:case_d2}, the algorithm designer outlines the solution strategy and provides the initial code. The bug fixer then generates a revised solution conditioned on the designer's output. Subsequently, a programming expert engages in a debate with the bug fixer to produce robust code that efficiently solves $q$, as depicted in Figure~\ref{fig:case_d3}.
Next, the project manager proposes a separate solution, which undergoes a detailed evaluation by the algorithm designer. Based on this critical feedback, the project manager refines the solution, as illustrated in Figure~\ref{fig:case_d4}.
Finally, a decision-maker reviews the collective contributions from all participants and delivers the final answer, as depicted in Figure~\ref{fig:case_d5}.
These interactions exemplify the iterative self-correction capability facilitated by the multi-type interactions.

\begin{algorithm*} % [H] 强制将算法放在当前位置
  \caption{Workflow of \textsc{TopoDIM}}
  \label{alg:example}
  \begin{algorithmic}[1] % [1] 表示显示行号
    % --- Input/Output ---
    \Require Initial graph $\mathcal{G}_\text{pri}$, query $\mathcal{Q}$
    \Ensure Communication topology $\hat{\mathcal{G}}$

    % --- 顺序语句 (Sequence) ---
    % \State Edge type $\mathcal{R} = \{\mathrm{conditioned,feedback,debate}\}$, $r_{constr} \in \{\mathrm{conditioned,debate}\}$ and $r_{free} = \mathrm{feedback}$

    % \State
    \noindent \note{// Stage 1: Centralized}

    \For {node i \textbf{in} $\{1,2,...,N\}$}
    \State $\mathbf{h}_i^{(0)} \gets \mathbf{e}_i^\text{role} + \mathbf{W}_q\phi(q)$
    \EndFor
    \For {query $q$ \textbf{in} $\mathcal{Q}$}

    \note{// Encoder}

    \State $\mathbf{h}^{l+1}_i \gets \sigma \big(\sum_{r\in\mathcal{R}_\text{pri}}\sum_{j\in\mathcal{N}_i^r}\frac{1}{c_{i,r}}\mathbf{W}_r^{(l)}\mathbf{h}_j^{(l)}+\mathbf{W}_0^{(l)}\mathbf{h}_i^{(l)}\big)$
    \State $\mathbf{H} \gets [\mathbf{h}^{l+1}_1,\mathbf{h}^{l+1}_2,...,\mathbf{h}^{l+1}_N]$

    \note{// Decoder}
    \State $p_\theta(\mathcal{G}\mid\mathbf{H})=\prod_{j=2}^N\prod_{i=1}^{j-1}p_\theta(r_{ij}\mid\mathbf{H},\mathcal{G}_{<ij})$
    \State $r_{ij}\gets \operatorname*{min}\{r\::\:F(r)\:\geq\:u\}$,
    $F(r)\gets\sum_{r^{\prime}=0}^rp_\theta(r')$

    \note{// Structural constraints}
    \State $p(r_{ij}\mid\cdot)\leftarrow p(r_{ij}\mid\cdot)\odot\mathbf{M}_{ij}$

    \note{// Sparsification}
    \State $\mathcal{E}_{\mathrm{final}}=\mathrm{TopK}\left(\{p(r_{ij})\}_{\forall i,j},1-\alpha\right)$
    \State $\mathcal{V}_{\mathrm{final}}=\{v\mid\exists(h,t,r)\in\mathcal{E}_{\mathrm{final}},v\in\{h,t\}\}$
    \State $\mathcal{G}_{\mathrm{final}} \gets (\mathcal{V}_{\mathrm{final}},\mathcal{E}_{\mathrm{final}},r_{\mathrm{final}})$
    \For {node i \textbf{in} $\psi(\mathcal{G_{\mathrm{final}}})$} \note{  // Breadth-first traversal}
    \State $\mathcal{S}_i \gets v_i(\mathcal{P})$
    \EndFor
    \State $a\leftarrow\text{Aggregate}(\{\mathcal{S}\}_{i=1}^N)$ \note{ // No multi-rounds dialogues}

    \State $\theta^{q+1}\leftarrow\theta^{q}-\delta\nabla_{\theta^{q}}J(\theta)$ \note{ // $\delta$ represents the learning rate during centralized training}
    \EndFor

    \noindent \note{// Stage 2: Decentralized}
    \For {query $q$ \textbf{in} $\mathcal{Q}$}
    \State $\mathcal{L}_{\mathrm{distill}}(\theta^\prime) \gets \sum_{v_i\in\mathcal{V}}\sum_{v_j\in\mathcal{V}\setminus v_i} D_{KL} \big( \pi_\theta(r_{ij} | \mathcal{G}_\text{pri}, q) \,||\, \pi_{\theta^\prime}(r_{ij} | \mathbf{h}_i^{(0)}, \mathbf{h}_j^{(0)}) \big)$ \note{  // No token overhead}
    \State $\theta^{\prime q+1}\leftarrow\theta^{\prime q}-\beta\nabla_{\theta^{\prime q}}\mathcal{L}_{\mathrm{distill}}$ \note{ // $\beta$ represents the learning rate during decentralized training}
    \EndFor
  \end{algorithmic}
  \label{Algo:topodim}
\end{algorithm*}

% This is an appendix.
\newpage

\end{document}